\documentclass[aps,prd,12point,twocolumn,nofootinbib,showpacs,superscriptaddress]{revtex4-2}

\usepackage{graphicx}
\usepackage{dcolumn}
\usepackage{tabu}
\usepackage{amsmath,amssymb}
\usepackage{float}
\usepackage{natbib}
\usepackage{balance}
\usepackage{braket}
\usepackage{mathrsfs}
\usepackage{bm}
\usepackage{lipsum}
\usepackage{comment}
\usepackage{xfrac}
\usepackage[dvipsnames]{xcolor}
\usepackage[title]{appendix}
\usepackage{txfonts}
\usepackage{xcolor}
\usepackage{slashed}
\usepackage[normalem]{ulem}
\usepackage[breaklinks=true,pdfborder={0 0 0},bookmarksopen]{hyperref}
\hypersetup{
	colorlinks   = true, 
	urlcolor     = magenta, 
	linkcolor    = blue, 
	citecolor   = Red 
}

\begin{document}

\title{Plane-wave and wavepacket neutrino flavor oscillations in vacuum in conformal coupling models}

\author{Fay\c{c}al Hammad}
\email{fhammad@ubishops.ca}
\affiliation{Department of Physics \& Astronomy,
Bishop's University,\\
2600 College St., Sherbrooke, QC, J1M 1Z7, Canada}
\affiliation{Physics Department, Champlain College-Lennoxville,\\
2580 College St., Sherbrooke, QC J1M 0C8, Canada}

\author{Parvaneh Sadeghi}
\email{psadeghi20@ubishops.ca}
\affiliation{Department of Physics \& Astronomy,
Bishop's University,\\
2600 College St., Sherbrooke, QC, J1M 1Z7, Canada}

\author{Nicolas Fleury}
\email{nfleury22@ubishops.ca}
\affiliation{Department of Physics \& Astronomy,
Bishop's University,\\
2600 College St., Sherbrooke, QC, J1M 1Z7, Canada}

\begin{abstract}
We investigate neutrino flavor oscillations in vacuum within a general conformal coupling model. We first examine the flavor oscillations within the plane-wave description of neutrinos, and then we extend our analysis to the wavepacket-based formalism. In both cases, we derive the general formulas for the flavor transition probability in arbitrary static and spherically symmetric spacetimes. We thoroughly discuss and assess in both cases the different possible ways ---\,dictated by the presence of the conformal coupling\,--- of computing the flavor transition probability. We show that the conformal invariance of the Dirac equation implies that the effect of conformal coupling on neutrino flavor oscillations modifies the oscillation length, but preserves the coherence length and unitarity. A detailed application to two-flavor neutrinos is then made and a numerical analysis is conducted within the well-known chameleon conformal coupling model.
\end{abstract}

\maketitle
\section{Introduction}\label{sec:Intro}
Attempts to explain the dark matter and dark energy problems of modern cosmology \cite{DarkReview} have led to various proposals for modifying the Einstein-Hilbert gravitational action. Such proposals can be classified into two main categories. In the first, one tweaks the gravitational action by putting in extra invariant curvature terms besides the Ricci scalar $R$ without adding any extra fields. In the second, one inserts into the action extra fields and geometric degrees of freedom in addition to possible extra invariant curvature terms. The latter models are known as scalar-tensor theories of gravity. Among the various proposals in the latter category are the conformal coupling models in which an extra field added to the action is conformally coupled to the matter sector of the action. The models that attracted much attention are the so-called chameleon model \cite{Chameleon} and the symmetron model \cite{Symmetron} as they exhibit a screening of the unwanted fifth force induced by the scalar field on matter. Although screening is also realized via other mechanisms, such as the Vainshtein mechanism \cite{Vainshtein, IntroVainshtein}, and within dilaton models \cite{Dilaton}, we focus in this paper on the chameleon- and symmetron-type models as they are the most studied in the literature \cite{ChameleonTests}. 

The gravitational sector in those conformal coupling models contains only the Einstein-Hilbert action and the usual action of a scalar field $\phi(x)$ with a kinetic term and a potential term. However, the matter sector of the models displays a conformal coupling between the gravitational field and the matter fields $\Psi_i(x)$. Such a gravitational coupling is said to be conformal because it amounts to replacing the curved-spacetime matter Lagrangian $\mathcal{L}_m(\Psi_i,g_{\mu\nu})$ by the Lagrangian $\mathcal{L}_m(\Psi_i,\tilde{g}_{\mu\nu})$. The metric $\tilde{g}_{\mu\nu}$ is related to the spacetime metric $g_{\mu\nu}$ of the gravitational sector by a Weyl conformal transformation \cite{Wald} $\tilde{g}_{\mu\nu}=\Omega^2(\phi)g_{\mu\nu}$, where the nonvanishing conformal factor $\Omega(\phi)$ is a functional of the scalar field $\phi(x)$. In the chameleon model the conformal factor has the form $\Omega(\phi)=\exp(\beta\, \phi/M_p)$, where $M_p$ stands for the reduced Planck mass and $\beta$ is an arbitrary dimensionless constant \cite{Chameleon}. In the symmetron model the conformal factor has the form $\Omega(\phi)=1+\frac{1}{2}\phi^2/M^2$ \cite{Symmetron}, where $M$ is an arbitrary mass scale. In both models, one assumes, for simplicity,  a universal conformal coupling in the sense that the scalar field $\phi$ couples identically to all matter fields. For more generality, however, one may introduce a different functional $\Omega_i(\phi)$ and/or a different scalar field $\phi_i$ for each matter field $\Psi_i$. Indeed, this would offer one extra possibility for violating the Equivalence Principle \cite{Will2018}.

On the other hand, the well-known theoretical proposal for neutrino flavor transition \cite{Pontecorvo1,Pontecorvo2} and its experimental discovery \cite{HistoryNO0, HistoryNO1,HistoryNO2,HistoryNO3,HistoryNO4,HistoryNO5,HistoryNO6,HistoryNO7,HistoryNO8,HistoryNO9,HistoryNO10,HistoryNO11} have given rise to a wide variety of scenarios for the origin (and even possible variation) of neutrino masses \cite{PhenomenologyWithN}. As the masses of neutrinos are the ones responsible for giving rise to the observed neutrino flavor oscillations, any possible variation of the masses could be confirmed by measuring neutrino oscillations \cite{PRL2004,PRL2005,NPB2005,JHP}. However, among the various proposed mechanisms that are of interest to us here are those that lead to varying masses for neutrinos based on the interaction of the latter with a scalar field \cite{JCAP2004,PRD2003,PLB2007,PRD2017,JCAP2018,DarkU2020} and, more specifically, the conformal coupling scenario. The effect of this kind of possible variation of neutrinos' masses on neutrino oscillations has more recently been examined by various authors \cite{PRD2021a,AddedRef,PRD2021b,DarkU2021,Arxiv2022}.

Now, it is also well known that the Dirac equation (and hence the fermion Lagrangian) in curved spacetime is conformally invariant \cite{QFT2}. Therefore, replacing the spacetime metric $g_{\mu\nu}$ by the conformal metric $\tilde{g}_{\mu\nu}$ inside the matter Lagrangian simply amounts to using the original metric $g_{\mu\nu}$ after replacing the fermion's mass $m$, its energy $E$ and its wavefunction $\psi$ by a new mass $\tilde{m}$, a new energy $\tilde{E}$ and a new wavefunction $\tilde{\psi}$. Thus, when a free neutrino mass eigenstate of mass $m_j$ and of energy $E_j$ is conformally coupled to the spacetime, it actually propagates in the curved spacetime of metric $g_{\mu\nu}=\Omega^{-2}\tilde{g}_{\mu\nu}$ as a `free' neutrino mass eigenstate of mass $\tilde{m}_j$ and of energy $\tilde{E}_j$. Under such a conformal deformation of the metric, the masses and energies of quantum particles are related, as usual \cite{What?}, by $\tilde{m}_j=\Omega m_j$ and $\tilde{E}_j=\Omega E_j$, respectively.

Therefore, since the characteristic length $L_{\rm osc}^{jk}$ of neutrino flavor oscillations is $4\pi\bar{E}/|m_j^2-m_k^2|$ (where $\bar{E}$ is the average energy of the neutrinos and $j$ and $k$ stand for the two mass eigenstates\footnote{See, e.g., Ref.\,\cite{NBook} for a rich introduction to neutrino oscillations and for many relevant references.}), we expect that the conformal coupling would simply lead to a new oscillation length $\tilde{L}_{\rm osc}^{jk}$ given by $4\pi\bar{\tilde{E}}/|\tilde{m}_j^2-\tilde{m}_k^2|$. In other words, we expect that a conformal coupling of neutrinos would lead to a new oscillation length which is roughly of the form $\tilde{L}_{\rm osc}^{jk}\sim L_{\rm osc}^{jk}/\Omega$. Given that the oscillation phase without conformal coupling is proportional to $L_p/L_{\rm osc}$ \cite{NBook}, where $L_p$ is the proper distance traveled by the conformally coupled neutrinos, the expected effect of the conformal coupling is then to induce oscillation phases that are proportional to the conformal factor.

The previous reasoning applies to neutrinos viewed as plane waves of definite wavelengths. Nevertheless, neutrino flavor oscillations are also very well studied in the literature using the wavepacket formalism as required by the uncertainties associated with the neutrinos production and detection processes \cite{NBook}. In Refs.\,\cite{NWPG,CV}, the effect of curved spacetime on the flavor oscillations within the wavepacket formalism has been investigated in much detail and shown to give rise to an oscillation phase that contains the ratio $I/L_{\rm osc}^{jk}$, where $I$ is an integral involving the components of the metric. In addition, however, the oscillation phase contains extra metric-dependent terms, as well as a damping term that also depends in a nontrivial way on the metric components. We therefore expect that while the plane-wave formalism might lead to simple expressions for the effect of a conformal coupling on neutrino oscillations, the wavepacket formalism could potentially display a more involved expression for the effect. Furthermore, the wavepacket approach in curved spacetime brings with it an additional complication due to the two possible ways of dealing with the propagation of the mass eigenstates wavepackets \cite{NWPG}.

Our first goal in this paper is to make the above hand-waving reasoning more precise by systematically deriving the flavor transition probability for conformally coupled neutrinos within both the plane-wave approach and the wavepacket approach. Our second goal is to give a physical interpretation of the subtleties brought in by the conformal coupling when dealing with wavepackets.

The remainder of this paper is organized as follows. In Sec.\,\ref{Sec:PW}, we derive the flavor transition probability for conformally coupled neutrinos within the plane-wave formalism. In Sec.\,\ref{Sec:WP}, we briefly recall the wavepacket approach to neutrino oscillations and the various subtleties it conceals, and then apply the approach to conformally coupled neutrinos. In Sec.\,\ref{Sec:Twoflavor}, we apply to the case of two-flavor neutrinos the general transition probabilities obtained in sections \ref{Sec:PW} and \ref{Sec:WP}. Finally, although we consider throughout this paper the case of universal conformal couplings, Sec.\,\ref{Sec:AppB} is devoted to discussing the way the results of sections \ref{Sec:PW} and \ref{Sec:WP} generalize further to the case of non-universal conformal couplings. We conclude this paper with a brief section in which we summarise and discuss our various results.
\section{Conformally coupled plane-wave neutrinos}\label{Sec:PW}
The curved-spacetime Dirac equation describing the dynamics of a spinor wavefunction $\psi_j$ associated to a neutrino mass eigenstate of mass $m_j$ reads $(i\gamma^\mu\nabla_\mu-m_j)\,\psi_j=0$\footnote{We set throughout the paper $\hbar=c=1$, and we work with the metric signature $(-,+,+,+)$. The Latin letters $(a,b)$ denote flat-spacetime indices, the letters $(j,k)$ denote neutrino mass eigenstates, while Greek letters $(\mu,\nu)$ denote curved-spacetime indices and the letters $(\alpha,\beta)$ will be used to denote neutrino flavors.}. This equation is obtained from the flat-spacetime Dirac equation by replacing the gamma matrices $\gamma^a$ by the curved-spacetime gamma matrices $\gamma^\mu=e^\mu_a\gamma^a$ and the partial derivatives $\partial_\mu$ by the spin covariant derivatives $\nabla_\mu=\partial_\mu+\frac{1}{8}\,\omega_\mu^{\,ab}[\gamma_a,\gamma_b]$. The spin connection $\omega_\mu^{\,ab}$ is related to the vierbeins $e^a_\mu$ and the Christoffel symbols $\Gamma_{\mu\nu}^\lambda$ by $\omega_\mu^{\,\,ab}=e^a_\nu\partial_\mu e^{\nu b}+\Gamma_{\mu\nu}^\lambda e^{a}_\lambda e^{\nu b}$. This curved-spacetime Dirac equation can be obtained from the fermion-matter Lagrangian \cite{QFT1} $\int{\rm d}^4x\sqrt{-g}\,\left[\tfrac{i}{2}\bar{\psi}_j\gamma^\mu\nabla_\mu\psi_j-\tfrac{i}{2}\nabla_\mu(\bar{\psi}_j)\gamma^\mu\psi_j-m_j\bar\psi_j\psi_j\right]$, where the metric determinant $\sqrt{-g}$ is that of the metric $g_{\mu\nu}$, and $\bar\psi$ is the adjoint spinor: $\bar\psi=\psi^\dagger\gamma^0$. However, for neutrinos conformally coupled to the metric via a scalar field $\phi$ their Lagrangian and their derived Dirac equation should both be written using the metric $\tilde{g}_{\mu\nu}=\Omega^2(\phi)g_{\mu\nu}$, the curved-spacetime gamma matrices $\tilde{\gamma}^\mu=\Omega(\phi)\gamma^\mu$ and the covariant derivatives $\tilde{\nabla}_\mu$ associated to the metric $\tilde{g}_{\mu\nu}$. We shall assume $\phi$ and $g_{\mu\nu}$ are time-independent.

As both the Lagrangian and the Dirac equation are conformally invariant, the conformally-coupled neutrinos' Lagrangian and their Dirac equation can be written in terms of the original spacetime metric $g_{\mu\nu}=\Omega^{-2}(\phi)\tilde{g}_{\mu\nu}$, respectively, as follows: 
\begin{align}\label{LagrangianDirac}
\!\!\!\!\!\mathcal{\tilde{L}}_m&=\!\int{\rm d}^4x\sqrt{-\tilde{g}}\left[\tfrac{i}{2}\bar{\psi}_j\tilde{\gamma}^\mu\tilde{\nabla}_\mu\psi_j-\tfrac{i}{2}\tilde{\nabla}_\mu(\bar{\psi}_j)\tilde{\gamma}^\mu\psi_j-m_j\bar\psi_j\psi_j\right]\nonumber\\
&=\int{\rm d}^4x\sqrt{-g}\!\left[\tfrac{i}{2}\tilde{\bar{\psi}}_j\gamma^\mu\nabla_\mu\tilde{\psi}_j-\tfrac{i}{2}\nabla_\mu(\tilde{\bar{\psi}}_j)\gamma^\mu\tilde{\psi}_j-\tilde{m}_j\tilde{\bar\psi}_j\tilde{\psi}_j\right],\!\!\!\!\\[5pt]
0&=(i\tilde{\gamma}^\mu\tilde{\nabla}_\mu-m_j)\psi_j=(i\gamma^\mu\nabla_\mu-\tilde{m}_j)\tilde{\psi}_j.
\end{align}
Since in the new expressions of both the Lagrangian and the Dirac equation the metric used is the original $g_{\mu\nu}$ of the gravitational sector, the scalar field $\phi$ is now hiding inside the new effective mass $\tilde{m}_j$ and the new wavefunction $\tilde{\psi}_j$. Indeed, these are given in terms of the original mass $m_j$ and the original wavefunction $\psi_j$, respectively, by \cite{QFT2,PRD2021a}:
\begin{equation}\label{ConformalMassPsi}
    \tilde{m}_j=\Omega(\phi)\,m_j, \qquad \tilde{\psi}_j(x)=\Omega^{\frac{3}{2}}(\phi)\,\psi_j(x).
\end{equation}
The new effective mass $\tilde{m}_j$ of each mass eigenstate becomes thus $\phi$-dependent and, hence, a position-dependent mass; i.e., a varying mass. Note that the conformal transformation of $\psi$ as given by Eq.\,(\ref{ConformalMassPsi}) is consistent with the requirement of preserving unitarity. Indeed, when working with the metric $\tilde{g}_{\mu\nu}$ the probability density to be integrated over a spatial region of volume $\tilde{V}$ of space is $\sqrt{\tilde{h}}\,\psi^\dagger\psi$, where $\tilde{h}$ is the determinant of the induced spatial metric in the region. On the other hand, when working with the metric $g_{\mu\nu}$, the probability density to be integrated over the spatial region of volume $V$ is the product $\sqrt{h}\,\tilde{\psi}^\dagger\tilde{\psi}$ which, according to the second identity in Eq.\,(\ref{ConformalMassPsi}) and the fact that $\sqrt{h}=\Omega^{-3}(\phi)\sqrt{\tilde{h}}$, is equal to $\sqrt{\tilde{h}}\,\psi^\dagger\psi$. In other words, the probability density is conserved by conformally transforming both the spacetime {\it and} the Dirac spinor.

Within the metric $g_{\mu\nu}$ the conformally coupled mass eigenstates satisfy the mass-shell condition $g_{\mu\nu}\tilde{p}_j^\mu\tilde{p}_j^\nu=-\tilde{m}_j^2$, where the four-momentum $\tilde{p}^\mu_j$ carries the effect of the scalar field $\phi$ through the following identity:
\begin{equation}\label{ConformalP}
    \tilde{p}_j^\mu=\tilde{m}_j\frac{{\rm d}x^\mu}{{\rm d}\tau}=\Omega(\phi)p_j^\mu,
\end{equation}
with $p_j^\mu=m_j{\rm d}x^\mu/{\rm d}\tau$ being the four-momentum that a neutrino mass eigenstate $\ket{\nu_j}$ of mass $m_j$ would have in the spacetime of metric $g_{\mu\nu}$ if it were not conformally coupled. 

Recall now that a neutrino mass eigenstate $\ket{\nu_j}$ {\it not} conformally coupled to spacetime, emitted at a point $(0,\vec{x}_A)$ in a spacetime of metric $g_{\mu\nu}$, is described at a detection point $(t,\vec{x}_B)$ of the spacetime by the following plane wave \cite{NBook}:
\begin{equation}\label{PlaneWave}
\ket{\nu_j^0(t,\vec{x}_B)}=\psi_j^0(t,\vec{x}_B)\ket{\nu_j}=e^{i\,\Phi_j^{\rm PW}}\ket{\nu_j},
\end{equation}
where the orthonormal basis states $\ket{\nu_j}$ of the Hilbert space satisfy $\braket{\nu_j|\nu_k}=\delta_{jk}$, and where the quantum phase $\Phi_j^{\rm PW}$ is evaluated using the Stodolsky prescription for computing the accumulated quantum phases of freely propagating quantum particles in curved spacetimes \cite{Stodolsky}:
\begin{equation}\label{AccumultatedPhase}
\Phi_j^{\rm PW}=-\int_A^B m_j\,{\rm d}\tau=\int_A^B g_{\mu\nu}p_j^\mu\,{\rm d}x^\nu.
\end{equation}
Note that we have attached here the superscript ``$0$'' to the mass eigenstate $\ket{\nu_j^0(t,\vec{x}_B)}$ and to its wavefunction $\psi_j^0(t,\vec{x}_B)$ in order to distinguish them from the corresponding quantities $\ket{\nu_j(t,\vec{x}_B)}$ and $\psi_j(t,\vec{x}_B)$ of the conformally coupled neutrinos propagating within the metric $\tilde{g}_{\mu\nu}$. Note also that we attached the superscript ``PW'' to the plane wave's quantum phase in order to distinguish it from the wavepacket's quantum phase $\Phi_j^{\rm WP}$ of Sec.\,\ref{Sec:WP}. Therefore, according to our discussion in the previous paragraph and according to Eq.\,(\ref{ConformalP}), a neutrino mass eigenstate $\ket{\nu_j}$ conformally coupled to the scalar field $\phi$ within a spacetime of metric $g_{\mu\nu}$ can be described at the point $(t,\vec{x}_B)$ by a plane wave $\ket{\tilde{\nu}_j(t,\vec{x}_B)}$ that reads
\begin{equation}\label{ConformalPlaneWave}
\ket{\tilde{\nu}_j(t,\vec{x}_B)}=\tilde{\psi}_j(t,\vec{x}_B)\ket{\nu_j}=e^{i\,\tilde{\Phi}_j^{\rm PW}}\ket{\nu_j},    
\end{equation}
where the quantum phase $\tilde{\Phi}_j^{\rm PW}$ is given by
\begin{equation}\label{ConformalAccumultatedPhase}
\tilde{\Phi}_j^{\rm PW}=-\int_A^B\tilde{m}_j\,{\rm d}\tau=\int_A^Bg_{\mu\nu}\tilde{p}_j^\mu\,{\rm d}x^\nu.
\end{equation}
Note that the orthonormal basis states of the Hilbert space in Eq.\,(\ref{ConformalPlaneWave}) satisfy again $\braket{\nu_j|\nu_k}=\delta_{jk}$. Before using the results (\ref{ConformalPlaneWave}) and (\ref{ConformalAccumultatedPhase}) to discuss the flavor transitions, we shall pause here to discuss the following important observation.

Our equation (\ref{ConformalPlaneWave}) differs from Eq.\,(30) of Ref.\,\cite{PRD2021a} mainly by the absence of any multiplicative factor made of powers of $\Omega(\phi)$. Such an absence in Eq.\,(\ref{ConformalPlaneWave}) might seem to be at odds with the second identity in Eq.\,(\ref{ConformalMassPsi}). One might indeed expect that the wavefunction in Eq.\,(\ref{ConformalPlaneWave}) would differ from the wavefunction in Eq.\,(\ref{PlaneWave}) only by the multiplicative factor $\Omega^{\frac{3}{2}}(\phi)$. One must bear in mind, however, that $\psi_j$ in Eq.\,(\ref{ConformalMassPsi}) is the wavefunction of the {\it conformally coupled} neutrino mass eigenstate as it propagates within the metric $\tilde{g}_{\mu\nu}$, whereas $\psi_j^0$ in our equation (\ref{PlaneWave}) is the wavefunction of a neutrino mass eigenstate {\it not} conformally coupled to the spacetime. There is no such a link between $\tilde{\psi}_j$ and $\psi_j^0$ as the one given by Eq.\,(\ref{ConformalMassPsi}) between $\tilde{\psi}_j$ and $\psi_j$. In fact, as we shall see shortly, the absence of any multiplicative factor made of powers of $\Omega(\phi)$ in our equation (\ref{ConformalPlaneWave}) is what will make the total flavor transition probability add up to unity, in contrast to what one would have obtained if multiplicative factors appeared in Eq.\,(\ref{ConformalPlaneWave}) as in Ref.\,\cite{PRD2021a}.

Now, a neutrino flavor $\ket{\nu_\alpha}$ is made out of a superposition of all three mass eigenstates $\ket{\nu_j}$ via the Pontecorvo-Maki-Nakagawa-Sakata (PMNS) unitary mixing matrix $U^*_{\alpha j}$ \cite{NBook}: $\ket{\nu_\alpha}=\sum_jU_{\alpha j}^*\ket{\nu_j}$. Therefore, the quantum mechanical probability for a conformally coupled neutrino produced at point $(0,\vec{x}_A)$ as an $\alpha$ flavor to be detected at the point $(t,\vec{x}_B)$ as a $\beta$ flavor is
\begin{align}\label{ConformalPlaneProba}
\mathcal{P}_{\alpha\rightarrow\beta}(t,\vec{x}_A,\vec{x}_B)&=|\braket{\nu_\beta|\sum_jU^*_{\alpha j}\tilde{\psi}_j(t,\vec{x}_B)|\nu_j}|^2\nonumber\\
&=\sum_{j,k}\,U^*_{\alpha j}U_{\beta j}U_{\alpha k} U^*_{\beta k}\,e^{i\,\tilde{\Phi}_{jk}^{\rm PW}}.
\end{align}
We have defined here, for convenience, the resulting phase difference by the usual notation $\tilde{\Phi}_{jk}^{\rm PW}\equiv\tilde{\Phi}_j^{\rm PW}-\tilde{\Phi}_k^{\rm PW}$. It must be noted here that although the wavefunction involved in Eq.\,(\ref{ConformalPlaneProba}) is the conformally transformed one, the mixing matrix $U_{\alpha j}^*$ is not affected by the conformal coupling. The reason is that the elements of such a matrix are dimensionless constants that specify only how much of each mass eigenstate in the Hilbert space is contained in the superposition that gives rise to a given flavor state. The conformal coupling of the mass eigenstates affects only their individual masses and wavefunctions living in the configuration space, it does not affect the superposition of the basis states $\ket{\nu_j}$. A more formal presentation of this argument is given in Appendix \ref{App}.

Eq.\,(\ref{ConformalPlaneProba}) is the general expression of the flavor transition probability of conformally coupled neutrinos propagating inside an arbitrary spacetime of metric $g_{\mu\nu}$. In order to use this result, we shall consider for simplicity in the remainder of this paper a radial propagation of neutrinos inside a static {\it and} spherically symmetric spacetime. Let the spacetime metric of the conformal coupling model be static and spherically symmetric, of the form
\begin{equation}\label{Metric}
{\rm d}s^2=-\mathcal{A}(r)\,{\rm d}t^2+\mathcal{B}(r)\,{\rm d}r^2+r^2\left({\rm d}\theta^2+\sin^2\theta\,{\rm d}\varphi^2\right).
\end{equation}
Here, the functions $\mathcal{A}(r)$ and $\mathcal{B}(r)$ are arbitrary functions of the radial coordinate $r$, which are everywhere regular except maybe at a singularity or on a horizon. 

Note that although according to Birkoff’s theorem one has $\mathcal{A}(r)\mathcal{B}(r)=1$ for most physical cases of such spacetimes, we shall assume throughout this paper the more general case $\mathcal{A}(r)\mathcal{B}(r)\neq1$ in view of its application to neutrino oscillations within modified gravity theories \cite{Buoninfante,NWPG} and in view of its application to neutrino oscillations within the gravitational field inside matter \cite{NWPG}. The usual case $\mathcal{A}(r)\mathcal{B}(r)=1$ of the exterior Schwarzschild solution found within general relativity will be dealt with in Sec.\,\ref{Sec:Twoflavor}.
\subsection{Integrating independently over $t$ and $r$}\label{Subsec:PWtrPhase}
Within a spacetime of metric (\ref{Metric}), the energy $\tilde{E}_j(\tilde{p})$ and momentum $\tilde{p}_j$ of a conformally coupled and radially propagating mass eigenstate of mass $\tilde{m}_j$ are, as perceived by an observer at infinity, given by the following expressions: 
\begin{equation}\label{InfinityObservables}
    \tilde{E}_j(\tilde{p})=-g_{00}\tilde{p}_j^0=\mathcal{A}(r)\,\tilde{m}_j\frac{{\rm d} t}{{\rm d}\tau},\qquad
    \tilde{p}_j=g_{rr}\tilde{p}_j^r=\mathcal{B}(r)\,\tilde{m}_j\frac{{\rm d}r}{{\rm d}\tau}.
\end{equation}
In the first step of each equality we used Eq.\,(\ref{ConformalMassPsi}), and in the second step we used Eq.\,(\ref{ConformalP}). The spacetime (\ref{Metric}) possesses a timelike Killing vector $K^\nu=(1,0,0,0)$, as well as a spacelike Killing vector $R^\mu=(0,0,0,1)$. The time-like Killing vector implies the conserved energies
\begin{equation}\label{ConservedEnergy}
E_j(p)=-g_{\mu\nu}K^\mu m_j\frac{{\rm d}x^\nu}{{\rm d}\tau}=\frac{\tilde{E}_j(\tilde{p})}{\Omega(\phi)},
\end{equation}
whereas the second Killing vector implies the conservation of the angular momentum which need not concern us here since we deal with radially propagating neutrinos. The conserved energy $E_j(p)$ is the energy an observer at infinity would perceive if the neutrino mass eigenstate of mass $m_j$ were freely propagating in the spacetime of metric $g_{\mu\nu}$. 

Combining the identities in Eq.\,(\ref{InfinityObservables}) with the mass-shell condition $-\tilde{m}_j^2=g_{\mu\nu}\tilde{p}_j^\mu \tilde{p}_j^\nu$, leads to
\begin{equation}\label{MassShell}
-\tilde{m}_j^2=-\frac{\tilde{E}_j^2(p)}{\mathcal{A}(r)}+\tilde{m}_j^2\mathcal{B}(r)\left(\frac{{\rm d}r}{{\rm d}\tau}\right)^2.
\end{equation}
On the other hand, plugging the second identity in Eq.\,(\ref{InfinityObservables}) into the mass-shell condition (\ref{MassShell}), we arrive at
\begin{align}\label{pinE(p)}
\tilde{p}_j&=\tilde{E}_j(\tilde{p})\sqrt{\frac{\mathcal{B}(r)}{\mathcal{A}(r)}}\left(1-\frac{\tilde{m}_j^2\mathcal{A}(r)}{\tilde{E}^2_j(\tilde{p})}\right)^{\frac{1}{2}}\nonumber\\
&\approx\sqrt{\frac{\mathcal{B}(r)}{\mathcal{A}(r)}}\left(\tilde{E}_j(\tilde{p})-\frac{\tilde{m}_j^2\mathcal{A}(r)}{2\tilde{E}_j(\tilde{p})}\right).
\end{align} 
In the second step, we expanded the expression up to the first order in the ratio $\tilde{m}_j^2/\tilde{E}_j(\tilde{p})$. Plugging Eq.\,(\ref{pinE(p)}) into Eq.\,(\ref{ConformalAccumultatedPhase}), and then integrating independently over $t$ and $r$, the accumulated phase of each mass eigenstate at the detection point $(t,\vec{x}_B)$ takes the following form:
\begin{align}\label{ComputedPlanePhase}
\tilde{\Phi}_{j}^{\rm PW}&=\int_A^B\left(-\tilde{E}_j(\tilde{p})\,{\rm d}t+\tilde{p}_j{\rm d}r\right)\nonumber\\
&=-E_j(p)\int_{A}^{B}\Omega(\phi)\,{\rm d}t+E_j(p)\int_{r_A}^{r_B}\Omega(\phi)\sqrt{\frac{\mathcal{B}(r)}{\mathcal{A}(r)}}\,{\rm d}r\nonumber\\
&\quad-\frac{m_j^2}{2E_j(p)}\int_{r_A}^{r_B}\Omega(\phi)\sqrt{\mathcal{A}(r)\mathcal{B}(r)}\,{\rm d}r\nonumber\\
&\equiv-E_j(p)\,\mathcal{T}+E_j(p)\,\mathcal{I}-\frac{m_j^2}{2E_j(p)}\,\mathcal{J}.
\end{align}
In the second step, we used identities (\ref{ConformalMassPsi}) and (\ref{ConservedEnergy}) that allowed us to move the conserved energy $E_j(p)$ and the constant mass $m_j$ out of the integrals, leaving behind the conformal factor $\Omega(\phi)$ that gets thus integrated along the path of the neutrinos from the emission to the detection point. In the third step, we introduced, for convenience, the notations $\mathcal{T}$, $\mathcal{I}$ and $\mathcal{J}$ for the three different integrals. The counterparts of these integrals in the case of neutrinos not conformally coupled are, respectively, the integrals $T$, $I_1$ and $I_2$ obtained in Ref.\,\cite{NWPG}. Inserting expression (\ref{ComputedPlanePhase}) of the phases into Eq.\,(\ref{ConformalPlaneProba}), we find the following $\mathcal{T}$-dependent transition probability from an $\alpha$-flavor neutrino to a $\beta$-flavor neutrino:
\begin{align}\label{TPlaneProbability}
&\mathcal{P}_{\alpha\rightarrow\beta}(\mathcal{T})=\sum_{j,k}U^*_{\alpha j}U_{\beta j}U_{\alpha k} U^*_{\beta k}\exp\Bigg[-iE_{jk}(p)\mathcal{T}+iE_{jk}(p)\mathcal{I}\nonumber\\
&\qquad\qquad\qquad\quad-i\frac{m_{j}^2}{2E_j(p)}\mathcal{J}+i\frac{m_{k}^2}{2E_k(p)}\mathcal{J}\Bigg].
\end{align}
We have used here the notation $E_{jk}(p)$ to denote the difference $E_j(p)-E_k(p)$, as well as the usual notation $\Delta m_{jk}^2$ to denote the difference $m_j^2-m_k^2$. 

We have now to consider the time-integral $\mathcal{T}$ to be unobservable and compute the average probability by averaging expression (\ref{TPlaneProbability}) over $\mathcal{T}$. This option is what is usually adopted in the literature when computing the transition probability \cite{NBook}. The reason for doing so is that while the distance traveled by the neutrinos might be known precisely, the times of emission and detection of the neutrinos might not. Averaging the transition probability (\ref{TPlaneProbability}) by integrating out the unobserved quantity $\mathcal{T}$ gives rise to the delta function $\delta [E_j(p)-E_k(p)]$, which thus simply eliminates the term $iE_{jk}(p)\mathcal{I}$ inside the exponential and replaces $E_j(p)$ and $E_k(p)$ in the remaining two terms by the average energy $E_0$ the neutrinos would have if they were massless \cite{NBook}. This leads then to the following expression for the averaged transition probability:
\begin{equation}\label{PlaneProbability}
\mathcal{P}_{\alpha\rightarrow\beta}=\sum_{j,k}U^*_{\alpha j}U_{\beta j}U_{\alpha k} U^*_{\beta k}\exp\left(-2\pi i\frac{\mathcal{J}}{L_{\rm osc}^{jk}}\right),
\end{equation}
where the {\it algebraic} oscillation length $L_{\rm osc}^{jk}$ is given by
\begin{equation}\label{PWOscLength}
L_{\rm osc}^{jk}=\frac{4\pi E_0}{\Delta m_{jk}^2}.
\end{equation}
When $\Omega(\phi)$ is a constant equal to unity, the general probability expression  (\ref{PlaneProbability}) reduces to the usual result for the case of neutrinos propagating in a curved spacetime without being conformally coupled. However, for the case of neutrinos conformally coupled and propagating within a Schwarzschild spacetime or within a Minkowski spacetime, for which $\mathcal{A}(r)\mathcal{B}(r)=1$, our result (\ref{PlaneProbability}) does not reduce to what is found in Ref.\,\cite{PRD2021a}. 

It is also worth stressing here the following important observation. The result (\ref{PlaneProbability}) shows that the effect of the conformal coupling of neutrinos to the scalar field in screening models shows up as an interesting cumulative phenomenon that is able to take into account the variation of the scalar field with position. However, when the scalar field $\phi$ settles down to its minimum value $\phi_0$ {\it all along} the neutrinos path, the factor $\Omega(\phi)$ in integral $\mathcal{J}$ can be moved out of the integration symbol. This would then simply amount to keeping the same formulas for neutrino flavor oscillations in curved spacetimes in the absence of any conformal coupling, and to just redefine the squared masses $m^2_j$ by the effective ones $\Omega(\phi_0)m^2_j$. In other words, the fifth-force screening mechanism, for which such modified-gravity models were designed in the first place, also works for neutrino flavor oscillations. That is, within local dense environments such as our solar system, and as long as one relies on the neutrino masses inferred from local solar-system neutrino-detection experiments, neutrino flavor oscillation in vacuum would not provide any signature of the conformal coupling in that case. Only outer-galactic neutrinos and neutrinos that traveled inside matter could possibly provide such a signature, for then the position dependence of the scalar field becomes significant.

Now, the above procedure of averaging over time is amply justified in the absence of a conformal coupling of neutrinos. Indeed, the time-integral $\mathcal{T}$ just reduces then to the total unmeasured time $T$ \cite{NWPG}. However, when neutrinos interact with a scalar field that displays a non-negligible variation with position along the path of the neutrinos towards the detector, actually performing the integral $\mathcal{T}$ becomes ambiguous. In fact, in such a case one can neither move the factor $\Omega(\phi)$ out of the integral nor replace the factor $\Omega(\phi)$ by its average value over the neutrinos path. In addition, as we shall see in Sec.\,\ref{Sec:WP}, the presence of the scalar field in the phase factor of neutrinos wavepackets makes the averaging-over-time approach even more problematic. For a rigorous treatment of the effect of the scalar field, one should therefore take into account the path of the neutrinos by using the link between ${\rm d}t$ and ${\rm d}r$ that follows by combining Eqs.\,(\ref{InfinityObservables}) and (\ref{MassShell}). Showing how to properly implement such a procedure in detail is the subject of the next subsection.

\subsection{Converting the time integral into a spatial integral}\label{Subsec:rPWPhase}
To take into account the presence of $\phi$ inside the time integral $\mathcal{T}$ when computing the phase, we need to express the time element ${\rm d}t$ in that integral in terms of the radial line element ${\rm d}r$. Isolating ${\rm d}t/{\rm d}\tau$ in terms of $\tilde{E}_j(\tilde{p})$ from the first identity in Eq.\,(\ref{InfinityObservables}) and then dividing the result by the ${\rm d}r/{\rm d}\tau$ that we isolate from Eq.\,(\ref{MassShell}), we find the following link between the coordinate elements ${\rm d}t$ and ${\rm d}r$ along the path of the $j$-th mass eigenstate: 
\begin{equation}\label{dr/dt}
{\rm d}t=\sqrt{\frac{\mathcal{B}(r)}{\mathcal{A}(r)}}\left(1-\frac{\tilde{m}_j^2\mathcal{A}(r)}{\tilde{E}_j^2(\tilde{p})}\right)^{-\frac{1}{2}}{\rm d}r\approx\sqrt{\frac{\mathcal{B}(r)}{\mathcal{A}(r)}}\left(1+\frac{\tilde{m}_j^2\mathcal{A}(r)}{2\tilde{E}_j^2(\tilde{p})}\right){\rm d}r.
\end{equation}
In the second step we expanded the square root up to the first order in $\tilde{m}_j^2/\tilde{E}_j^2(\tilde{p})$. We need now to use this link to find the quantum phase of each mass eigenstate. However, to achieve that purpose we cannot merely insert Eq.\,(\ref{dr/dt}) into the time integral in Eq.\,(\ref{ComputedPlanePhase}) corresponding to the $j$-th mass eigenstate and then merely replace everywhere $j$ by $k$ for the case of the $k$-th mass eigenstate. The reason is that a neutrino is detected as having a certain flavor when the mass eigenstates making that flavor are {\it all} detected at the same point in space and at the {\it same time}. Therefore, since the detection position is $r=r_B$ for all the mass eigenstates, a common detection time is obtained only if one turns all time integrals into spatial integrals using a {\it unique} link between the elements ${\rm d}t$ and ${\rm d}r$. The time integral that needs to be performed in Eq.\,(\ref{ComputedPlanePhase}) for any mass eigenstate should then be the time integral corresponding to the {\it most} massive state of the three mass eigenstates, for the wave front of the most massive mass eigenstate is the last one to arrive at $r=r_B$. 

Let us denote the mass and energy of the most massive state by $\tilde{m}_*$ and $\tilde{E}_*$, respectively. Then the time integral that should be inserted into the calculation (\ref{ComputedPlanePhase}) of the phase for all the mass eigenstates is
\begin{align}\label{MassiveTime}
    \mathcal{T}_*&=\int_A^B\Omega(\phi){\rm d}t=\int_{r_A}^{r_B}\Omega(\phi)\sqrt{\frac{\mathcal{B}(r)}{\mathcal{A}(r)}}\left(1+\frac{\tilde{m}_*^2\mathcal{A}(r)}{2\tilde{E}_*^2(\tilde{p})}\right){\rm d}r\nonumber\\
    &=\mathcal{I}+\frac{m_*^2}{2E_*^2(p)}\mathcal{J}.
\end{align}
Inserting this time $\mathcal{T}_*$ into the second line of Eq.\,(\ref{ComputedPlanePhase}), the resulting quantum mechanical phase of the $j$-th mass eigenstate is given by the following expression
\begin{equation}\label{trPhase}
    \tilde{\Phi}_j^{\rm PW}=-\left[\frac{E_j(p)m_*^2}{2E_*^2(p)}+\frac{m_j^2}{2E_j(p)}\right]\mathcal{J}.
\end{equation}
Therefore, inserting into expression (\ref{ConformalPlaneProba}) of the transition probability the phase difference $\tilde{\Phi}^{\rm PW}_{jk}$ one obtains by subtracting from Eq.\,(\ref{trPhase}) the corresponding expression for the $k$-th mass eigenstate, the transition probability takes the following form
\begin{align}\label{MassiveBasedProbability}
\mathcal{P}_{\alpha\rightarrow\beta}&=\sum_{j,k}U^*_{\alpha j}U_{\beta j}U_{\alpha k} U^*_{\beta k}\nonumber\\
&\qquad\times\exp\left(-i\left[\frac{m_j^2}{2E_j(p)}-\frac{m_{k}^2}{2E_k(p)}+\frac{E_{jk}(p)m_{*}^2}{2E_*^2(p)}\right]\mathcal{J}\right).
\end{align}
Here, $E_{jk}(p)$ stands again for the difference $E_j(p)-E_k(p)$. Note that there is no more time-dependence in this expression of the probability, and therefore no averaging over time needs to be performed within this approach.

Next, approximating the conserved energies $E_j(p)$ and $E_k(p)$ of each mass eigenstate by the common average energy $E_0$ of massless neutrinos, the transition probability (\ref{MassiveBasedProbability}) takes exactly the same form as the one obtained within the previous approach; i.e., it takes the form given by Eq.\,(\ref{PlaneProbability}) in which the algebraic oscillation length $L_{\rm osc}^{jk}$ is again given by Eq.\,(\ref{PWOscLength}). Note that one may go up to the first order in $m^2/E$ and approximate the conserved energies $E_j(p)$ and $E_k(p)$ by writing $E_{j}(p)\approx E_0+\xi m_j^2/2E_0$ and $E_{k}(p)\approx E_0+\xi m_k^2/2E_0$, respectively. The dimensionless constant $\xi$ is smaller than one and it entirely depends on the particulars of the neutrino production process, such as the characteristics of the interaction and the nature of the other particles taking part in the process. Within the wavepacket treatment of the neutrino production process, for instance, the constant $\xi$ can be fully expressed in terms of the dispersion width $\sigma_P$ of the produced neutrino's wavepacket and the corresponding group velocity of the latter \cite{Giunti}. By expanding these conserved energies up to the first order in $m^2/E$, however, one would only end up with correction terms that are of the second order in $m^2/E$ in Eqs.\,(\ref{PlaneProbability}) and (\ref{PWOscLength}) where such terms have already been discarded. Therefore, both approaches are consistent and lead to the same final result. Furthermore, in the limit $\Omega(\phi)\rightarrow1$ our results reduce to the ones obtained in previous works on neutrinos not conformally coupled \cite{NBook,NWPG}. 

It is also worth noting here the following. If we had not taken into account the fact that all the mass eigenstates are detected at the same time, instead of the common time integral $\mathcal{T}_*$ given by Eq.\,(\ref{MassiveTime}), we would have inserted into each phase $\tilde{\Phi}_j^{\rm PW}$ the time integral $\mathcal{T}_j$ obtained by using the corresponding link (\ref{dr/dt}) between ${\rm d}t$ and ${\rm d}r$. This would have led us to the same issue already discussed by various authors in the literature, which gave rise to a controversy over the presence or absence of a multiplicative factor of 2 in the oscillation length (\ref{PWOscLength}) \cite{Bhattacharya,Fornengo,Lipkin,HalfGiunti,Zhang,Ren,Chakraborty1,Chakraborty2}. It should be noted here that the correct and more rigorous derivation of the gravitationally induced neutrino flavor oscillation was first conducted in Ref.\,\cite{Bhattacharya} for neutrinos not conformally coupled. The latter study has more recently been elaborated further and applied in Ref.\,\cite{1805.01874} to study flavor interference patterns of neutrinos scattering off a static black hole by combining both analytic and numerical methods which allows one to include even non-radially propagating neutrinos.

\section{Conformally coupled wavepacket neutrinos}\label{Sec:WP}
The wavepacket approach to neutrino flavor oscillations in the absence of conformal coupling aims at taking into account the uncertainties associated with the production and detection processes \cite{NBook}. It consists in replacing the plane waves $\ket{\nu^0_j(t,x)}$ of each mass eigenstate $j$ by a superposition (usually chosen to be a Gaussian) of plane waves of infinitely many possible values of the momentum $p_j$, all centered on a mean value $\bar{p}_j$. Thus, in the wavepacket approach the normalized wavefunctions of the produced and the detected neutrinos at any spacetime point $(t,\vec{x})$ are taken to be
\begin{equation}\label{CurvedWavePacket}
\psi^0_j(t,\vec{x})=\int_{-\infty}^{+\infty}\frac{{\rm d}^3{\bf p}_j}{(2\pi)^3}\left(\frac{2\pi}{\sigma_{p}^2}\right)^{\frac{3}{4}} e^{-\frac{({\bf p}_j-\bar{\bf p}_j)^2}{4\sigma_{p}^{2}}}e^{ip_{j\mu}x^\mu}.
\end{equation}
The Gaussian dispersion width $\sigma_{p}$ should be distinguished for produced and detected neutrinos by introducing two different widths $\sigma_{pP}$ and $\sigma_{pD}$, respectively \cite{NBook}. However, for the sake of clarity and simplicity, in what follows we only take into consideration one dispersion width $\sigma_p$, to be associated with the production process. Therefore, we assume that the detected mass eigenstate state $\ket{\nu_k}$ is just one of the orthonormal basis states of the Hilbert space. Note, also, that we do not consider a covariant Gaussian wavepacket, the exponential factor of which would have been $\exp[-(p_j-\bar{p}_j)^\mu(p_j-\bar{p}_j)_\mu/(4\sigma_p)^{2}]$ \cite{Naumov}. The reason is that, doing so, would only bring corrections to our results that are of orders higher than $\mathcal{O}(m_j^4/E_j^2)$, whereas we restrict ourselves here only to the first order in $m_j^2/E_j$. 

As our goal here is to examine only radially propagating neutrinos, we shall consider again the static and spherically symmetric metric (\ref{Metric}), and restrict our study to the one-dimensional version of the wavepacket (\ref{CurvedWavePacket}).  For radially propagating conformally coupled neutrinos within a spacetime of metric $g_{\mu\nu}$, the Gaussian wavepacket (\ref{CurvedWavePacket}) takes the following form in one dimension:
\begin{equation}\label{1DCurvedWavePacket}
\tilde{\psi}_j(t,r)=\int_{-\infty}^{+\infty}\frac{{\rm d}\tilde{p}_j}{2\pi}\left(\frac{2\pi}{\tilde{\sigma}_{p}^2}\right)^{\frac{1}{4}} e^{-\frac{(\tilde{p}_j-\bar{\tilde{p}}_j)^2}{4\tilde{\sigma}_{p}^{2}}}e^{-i\tilde{E}_jt+i\tilde{p}_jr}.
\end{equation}
Now, the subtlety we alluded to in the Introduction concerning the wavepacket approach resides in the two possible ways of computing the accumulated phase of the wavepacket (\ref{1DCurvedWavePacket}) when working in curved spacetimes. It was shown in Ref.\,\cite{NWPG} that to properly implement the wavepacket approach in curved spacetimes, one needs to first locally evaluate the Gaussian integral (\ref{1DCurvedWavePacket}) and then compute its accumulated phase at the detection point. For this reason, we shall compute the transition probability using this approach, but for the sake of completeness we shall discuss in detail in subsection \ref{Sec:WPAfter} why a second approach for dealing with the wavepackets leads one further astray when working with conformally coupled neutrinos.

The first approach consists then of building the Gaussian wavepacket by first evaluating integral (\ref{1DCurvedWavePacket}) for all values of $t$ and $r$. To carry out the integration, we need to express the energy $\tilde{E}_j(\tilde{p})$ in the exponential explicitly in terms of the momentum $\tilde{p}_j$. Since the Gaussian factor is rapidly suppressed away from the mean momentum $\bar{\tilde{p}}_j$, we may, to a good approximation, just expand $\tilde{E}_j(\tilde{p})$ in terms of $\tilde{p}_j$ around $\bar{\tilde{p}}_j$ to a few powers of $(\tilde{p}_j-\bar{\tilde{p}}_j)$. As shown in Ref.\,\cite{NWPG}, in order to obtain the term responsible for the spreading over time of the wavepacket, one needs to expand $\tilde{E}_j(\tilde{p})$ at least up to the second power of $(\tilde{p}_j-\bar{\tilde{p}}_j)$. However, since the resulting corrections are of the order of $\tilde{m}^2_j/\tilde{E}^3_j(\tilde{p})$ and the resulting expressions are very involved even for the case of neutrinos not conformally coupled \cite{NWPG}, we shall restrict ourselves here to the linear term in the expansion of $\tilde{E}_j(\tilde{p})$. The effect of the conformal coupling of neutrinos on their flavor oscillations will indeed be clearly identified and more easily confronted to experiments even at this order. Let us therefore perform the following Taylor expansion of the energy:
\begin{align}\label{FlatEExpansion}
\tilde{E}_j(\tilde{p})&=\bar{\tilde{E}}_j+\frac{\partial \tilde{E}_j(\tilde{p})}{\partial \tilde{p}}\bigg|_{\tilde{p}=\bar{\tilde{p}}_j}\!\left(\tilde{p}_j-\bar{\tilde{p}}_j\right)\equiv\bar{\tilde{E}}_j+\tilde{\varv}_j\left(\tilde{p}_j-\bar{\tilde{p}}_j\right).
\end{align} 
Here, $\bar{\tilde{E}}_j$ represents an average energy for the wavepacket associated to the mass eigenstate $\ket{\nu_j}$, and $\tilde{\varv}_j$ represents the group velocity of the wavepacket. To evaluate the parameter $\tilde{\varv}_j(r)$ which is position-dependent, we first substitute the ${\rm d}r/{\rm d}\tau$ we extract from the second identity in Eq.\,(\ref{InfinityObservables}) into Eq.\,(\ref{MassShell}) to find, 
\begin{equation}\label{E(p)inp}
\tilde{E}_j(\tilde{p})=\sqrt{\frac{\mathcal{A}(r)}{\mathcal{B}(r)}}\left[\tilde{p}_j^2+\tilde{m}_j^2\mathcal{B}(r)\right]^{\frac{1}{2}}.
\end{equation}
From this identity, we then find the following expression for $\tilde{\varv}_j(r)$:
\begin{equation}\label{v}
    \tilde{\varv}_j(r)\approx\sqrt{\frac{\mathcal{A}(r)}{\mathcal{B}(r)}}\left(1-\frac{\tilde{m}_j^2\mathcal{B}(r)}{2\bar{\tilde{p}}_j^2}\right).
\end{equation}
Substituting this expression into Eq.\,(\ref{FlatEExpansion}), and then plugging the latter into the Gaussian integral (\ref{1DCurvedWavePacket}), we arrive at the following expression for the wavepacket at any spacetime position $(t,r)$:
\begin{equation}\label{Final1DCurvedWP}
    \tilde{\psi}_j(t,r)=\left(\frac{2\tilde{\sigma}_{p}^2}{\pi}\right)^{\frac{1}{4}}\exp\left[-i\bar{\tilde{E}}_jt+i\bar{\tilde{p}}_jr-\tilde{\sigma}_{p}^2(r-\tilde{\varv}_jt)^2\right].
\end{equation}

We need now to evaluate, using the Stodolsky prescription, the accumulated quantum mechanical phase of this wavepacket as it propagates towards the detector at $r=r_B$. For that purpose, we need to integrate over $t$ and $r$ {\it only} those terms that do contribute to the quantum phase \cite{NWPG}. 
On the other hand, just like in the case of plane waves studied in Sec.\,\ref{Sec:PW}, the average energy in the phase term $-i\bar{\tilde{E}}_jt$ is not the conserved quantity along the path of the wavepacket. The conserved average energy now is $\bar{E}_j=\bar{\tilde{E}}_j/\Omega(\phi)$, as imposed by Eq.\,(\ref{ConservedEnergy}). Therefore, when performing the time integral to get the accumulated quantum phase we should integrate $\Omega(\phi)\bar{E}_j$ by moving only $\bar{E}_j$ out of the integral and leaving inside the latter the conformal factor $\Omega(\phi)$. However, unlike in the plane wave case, doing so would give here two different times inside the wavepacket (\ref{Final1DCurvedWP}): the time $\mathcal{T}$ of Sec.\,\ref{Sec:PW} in the phase part of the exponential and a total time $T$ in the decaying part of the exponential. With two different times at hand, there would be no possibility for averaging the final transition probability over time as we did in Sec.\,\ref{Sec:PW}. 

There are two different possible ways of solving this issue. The first, consists of averaging the wavepacket (\ref{Final1DCurvedWP}) over the time $t$ {\it before} computing the quantum mechanical phase the wavepacket would accumulate as it propagates towards the detector.  The second way consists of computing {\it first} the final quantum phase of the wavepacket by integrating the phase part of expression (\ref{Final1DCurvedWP}) over $t$ and $r$ after taking into account an analog of relation (\ref{dr/dt}) between ${\rm d}t$ and ${\rm d}r$. We shall now examine both methods in detail and discuss their respective outcomes, starting with the first method.

\subsection{Averaging over $t$}
Averaging over time by integrating the Gaussian (\ref{Final1DCurvedWP}) over the coordinate time $t$, we easily find the following time-independent expression for the wavepacket at any radial position $r$:
\begin{equation}\label{tAveragedWP}
    \tilde{\psi}_j(r)\propto\left(\frac{2\pi}{\tilde{\sigma}_{p}^2\tilde{\varv}_j^4}\right)^{\frac{1}{4}}\exp\left[-i\left(\frac{\bar{\tilde{E}}_j}{\tilde{\varv}_j}-\bar{\tilde{p}}_j\right)r-\frac{\bar{\tilde{E}}_j}{4\tilde{\sigma}_p^2{\tilde{\varv}_j^2}}\right].
\end{equation}
Note that this result is valid only for a time-independent conformal factor $\Omega(\phi)$. To find the final value of this wavepacket at the detector, we need to compute the accumulated quantum mechanical phase {\it \`a la} Stodolsky by integrating the imaginary part in the exponential while simply evaluating the real term at $r=r_B$. To properly compute the quantum phase, we make use of Eqs.\,(\ref{pinE(p)}) and (\ref{v}) to find the following expression by keeping only terms up to the first order in $\tilde{m}_j^2/\bar{\tilde{E}}_j$:
\begin{align}\label{tAveragedWPPhase}
    \tilde{\Phi}_j^{\rm WP}&=-\int_{r_A}^{r_B}\frac{\tilde{m}^2_j}{2\bar{\tilde{E}}_j}\sqrt{\frac{\mathcal{B}^3(r)}{\mathcal{A}(r)}}{\rm d}r-\int_{r_A}^{r_B}\frac{\tilde{m}^2_j}{2\bar{\tilde{E}}_j}\sqrt{\mathcal{A}(r)\mathcal{B}(r)}{\rm d}r\nonumber\\
    &=-\frac{m_j^2}{2\bar{E}_j}\left(\int_{r_A}^{r_B}\!\Omega(\phi)\sqrt{\frac{\mathcal{B}^3(r)}{\mathcal{A}(r)}}{\rm d}r\!+\!\int_{r_A}^{r_B}\!\Omega(\phi)\sqrt{\mathcal{A}(r)\mathcal{B}(r)}{\rm d}r\right)\nonumber\\
    &\equiv-\frac{m_j^2}{2\bar{E}_j}\left(\mathcal{K}+\mathcal{J}\right).
\end{align}
In the second step we used Eqs.\,(\ref{ConformalMassPsi}) and (\ref{ConservedEnergy}), and in the third step we introduced the notation $\mathcal{K}$ for the new integral involving the conformal factor. With this accumulated phase at the detection point, the wavepacket (\ref{tAveragedWP}) leads to the following transition probability:
\begin{align}\label{ConformaltAvgWPProba}
\mathcal{P}_{\alpha\rightarrow\beta}(r_B)&=|\braket{\nu_\beta|\sum_jU^*_{\alpha j}\tilde{\psi}_j(r_B)|\nu_j}|^2\nonumber\\
&\propto\sum_{j,k}\,U^*_{\alpha j}U_{\beta j}U_{\alpha k} U^*_{\beta k}\nonumber\\
&\qquad\times\exp\left(-2\pi i\frac{\mathcal{J}+\mathcal{K}}{L_{\rm osc}^{jk}}-\frac{E_0\,\Omega(\phi_B)}{4\sigma_p^2\varv_j^2}-\frac{E_0\,\Omega(\phi_B)}{4\sigma_p^2\varv_k^2}\right).
\end{align}
Here, the algebraic oscillation length $L_{\rm osc}^{jk}$ is again given by Eq.\,(\ref{PWOscLength}), whereas $\varv_j$ and $\varv_k$ together with $\Omega(\phi_B)$ are all evaluated at $r=r_B$. Note that the average energy $E_0$ of the massless-neutrinos approximation appearing in Eq.\,(\ref{PWOscLength}) comes here again from the approximation $\bar{E}_j=E_0+\xi m_j^2/2E_0$ that arises from a quantum field theoretical treatment of neutrino wavepackets \cite{Giunti}. The dimensionless constant $\xi$ being less than one \cite{Giunti}, we make here and in the next subsection the approximation $\bar{E}_j\approx E_0$ that restricts us to the first order in $m^2_j/E_0$. Note also that we have used the fact that $\tilde{\varv}_j=\varv_j$, $\tilde{\varv}_k=\varv_k$ and $\tilde{\sigma}_p^2=\sigma_p^2$. The conformal invariance of $\varv_j(r)$ can be seen by inserting Eq.\,(\ref{ConformalP}) and the first identity in Eq.\,(\ref{ConformalMassPsi}) into Eq.\,(\ref{v}). The conformal invariance of $\sigma_p^2=\tilde{g}_{\mu\nu}\sigma_p^\mu\sigma_p^\nu$ stems from the fact that $\tilde{g}_{\mu\nu}=\Omega^2(\phi)g_{\mu\nu}$ and the fact that $\tilde{\sigma}_p^\mu=\Omega(\phi)\sigma_p$, as it follows from Eq.\,(\ref{ConformalP}); leading to $\tilde{\sigma}_p^2=g_{\mu\nu}\tilde{\sigma}_p^\mu\tilde{\sigma}_p^\nu=\sigma_p^2$.

We clearly see now that in the limit $\Omega(\phi)\rightarrow1$, the result (\ref{ConformaltAvgWPProba}) does not reduce to what is found for the case of neutrinos not conformally coupled. This shows that this approach is not valid.
\subsection{Averaging over $T$}
Because of the presence of the conformal factor $\Omega(\phi)$, we compute here the accumulated quantum mechanical phase of the wavepacket (\ref{Final1DCurvedWP}) by turning the time integral of the phase term into a spatial integral thanks to the link (\ref{dr/dt}). According to our discussion in subsection \ref{Subsec:rPWPhase}, we should here take care again of inserting the analog of the unique time integral (\ref{MassiveTime}) of the most massive mass eigenstate into the quantum phase of the wavepacket of each mass eigenstate. The analog of the integral (\ref{MassiveTime}) for wavepackets is obtained by replacing $E_*(p)$ in Eq.\,(\ref{MassiveTime}) by the average energy $\bar{E}_*$ of the wavepacket of the most massive state. Therefore, the expression of the wavepacket of the $j$-th mass eigenstate at the detection point $r=r_B$, such that $r_B-r_A=L$, takes the following form
\begin{equation}\label{Final2DCurvedWP}
    \tilde{\psi}_j(T,L)=\left(\frac{2\sigma_{p}^2}{\pi}\right)^{\frac{1}{4}}\exp\left[-i\left(\frac{\bar{E}_jm_*^2}{2\bar{E}_*^2}+\frac{m_j^2}{2\bar{E}_j}\right)\mathcal{J}\!-\!\sigma_{p}^2(L\!-\!\varv_jT)^2\right].
\end{equation}
We have used here the conformal invariance of both $\sigma_p^2$ and $\varv_j$ as discussed in the previous subsection, and we have introduced the unmeasured total coordinate time $T=\int_A^B{\rm d}t$. Therefore, adopting again the zeroth-order approximation for the average energies, $\bar{E}_j=E_0+\mathcal{O}(m_j^2/E_0)$, the transition probability takes, up to the first order in $m^2_j/E_0$, the following form
\begin{align}\label{ConformalWPProba}
&\mathcal{P}_{\alpha\rightarrow\beta}(T,L)=|\braket{\nu_\beta|\sum_jU^*_{\alpha j}\,\tilde{\psi}_j(T,L)|\nu_j}|^2\nonumber\\
&=\sum_{j,k}\,U^*_{\alpha j}U_{\beta j}U_{\alpha k} U^*_{\beta k}\left(\frac{2\sigma_p^2}{\pi}\right)^{\frac{1}{2}}\nonumber\\
&\qquad\times\exp\left[-2\pi i\frac{\mathcal{J}}{L_{\rm osc}^{jk}}-\sigma_{p}^2(L-\varv_jT)^2-\sigma_{p}^2(L-\varv_kT)^2\right].
\end{align}
Note that, as in the case of the plane waves dealt with in Sec.\,\ref{Sec:PW}, there is no overall multiplicative factor involving $\Omega(\phi)$ or its powers. Furthermore, the oscillation length $L_{\rm osc}^{jk}$ in Eq.\,(\ref{ConformalWPProba}) is given again by the standard expression (\ref{PWOscLength}). Averaging now this probability over the unmeasured time $T$, we easily find the following final expression for the transition probability:
\begin{equation}\label{FinalConformalWPProba}
\mathcal{P}_{\alpha\rightarrow\beta}(L)=\sum_{j,k}\,U^*_{\alpha j}U_{\beta j}U_{\alpha k} U^*_{\beta k}\,\exp\left[-2\pi i\frac{\mathcal{J}}{L_{\rm osc}^{jk}}-\left(\frac{L}{L_{\rm coh}^{jk}}\right)^2\right].
\end{equation}
We have introduced here the algebraic coherence length\footnote{We would like to point out here the following possibly confusing typos in Ref.\,\cite{NWPG}: The displayed oscillation and coherence lengths in Ref.\,\cite{NWPG} are the positive physical ones as they carry the absolute value $|\Delta m_{jk}^2|$, whereas inside the exponential of all the transition probabilities displayed there, that term should have been displayed in its {\it algebraic} form: $\Delta m_{jk}^2$.}:
\begin{equation}\label{Coherence}
L_{\rm coh}^{jk}\!=\!\frac{\sqrt{\varv_j^2+\varv_k^2}}{\sigma_p(\varv_j-\varv_k)}\approx\frac{4\sqrt{2}\bar{E}^2\sigma_x}{\Delta m_{jk}^2}\mathcal{A}^{-1}(r_B)\left(1\!-\!\frac{m_j^2+m_k^2}{4\bar{E}^2}\mathcal{A}(r_B)\right).
\end{equation}
In the second step we used expression (\ref{v}) for the group velocities $\varv_j$ and $\varv_k$. This coherence length is nothing but the coherence length already found in Ref.\,\cite{NWPG} for neutrinos not conformally coupled. The quadratic term $(L/L_{\rm coh}^{jk})^2$ in Eq.\,(\ref{FinalConformalWPProba}) is the term responsible for damping the flavor oscillations with increasing distance due to the spreading of the wavepacket. 

Very important to note also, is the fact that this result shows that the conformal factor $\Omega(\phi)$ does not contribute to the damping of the flavor oscillations. The conformal coupling of neutrinos affects thus only the oscillation phase. Furthermore, since the result (\ref{FinalConformalWPProba}) remains valid no matter how slowly the field $\phi$ varies with position, it should remain valid for a constant factor $\Omega(\phi_0)$ achieved when the field $\phi_0$ settles down to its minimum value $\phi$ within dense environments. Therefore, we see that even the wavepacket approach to neutrino flavor oscillations indicates that the latter are also subject to a screening mechanism. Of course, the result (\ref{FinalConformalWPProba}) also remains valid in the limit $\Omega(\phi)\rightarrow1$, for which we simply recover the result obtained in Ref.\,\cite{NWPG} for the case of neutrinos not conformally coupled. 

\subsection{Building the wavepacket at the detector}\label{Sec:WPAfter}
The second approach for dealing with wavepackets in curved spacetimes consists of building the wavepacket of each mass eigenstate from plane waves only after the latter have all reached the detector \cite{CV}. Therefore, the Gaussian integral one needs to evaluate would be the following
\begin{equation}\label{WPAfter}
\tilde{\psi}_j(T,L)=\int_{-\infty}^{+\infty}\frac{{\rm d}\tilde{p}_j}{2\pi}\left(\frac{2\pi}{\tilde{\sigma}_{p}^2}\right)^{\frac{1}{4}} e^{-\frac{(\tilde{p}_j-\bar{\tilde{p}}_j)^2}{4\tilde{\sigma}_{p}^{2}}}e^{i\tilde{\Phi}_j^{\rm PW}}.
\end{equation}
Here, the final accumulated quantum phase $\tilde{\Phi}_j^{\rm PW}$ of each constituent plane wave is thus computed before building the wavepacket. We have now two different options again for computing $\tilde{\Phi}_j^{\rm PW}$. One may compute the quantum phase $\tilde{\Phi}_j^{\rm PW}$ either by integrating independently over $t$ and $r$ as done in subsection \ref{Subsec:PWtrPhase}, or by converting the time integral into a spatial integral as done in subsection \ref{Subsec:rPWPhase}. In the first case, one should insert into the Gaussian integral (\ref{WPAfter}) the quantum phase given by Eq.\,(\ref{ComputedPlanePhase}), whereas in the second case one should insert into the Gaussian (\ref{WPAfter}) the quantum phase given by Eq.\,(\ref{trPhase}). We clearly see now the big issue one faces when adopting this approach. Both phases (\ref{ComputedPlanePhase}) and (\ref{trPhase}) would prevent one from properly performing the Gaussian integral (\ref{WPAfter}). Indeed, the integration variable in that Gaussian is the momentum $\tilde{p}_j$ in which the implicit conformal factor $\Omega(\phi)$ should now be evaluated at $r=r_B$, whereas both phases (\ref{ComputedPlanePhase}) and (\ref{trPhase}) contain through the energies $E_j(p)$ and $E_*(p)$ a momentum $p_j$ that is {\it decoupled} from the conformal factor $\Omega(\phi)$ as the latter has already been integrated over the path of the neutrinos inside integrals $\mathcal{T}$, $\mathcal{I}$ and $\mathcal{J}$. This shows that this approach is also inadequate when applied to conformally coupled neutrinos.

\section{Two-flavor neutrinos case}\label{Sec:Twoflavor}
All the results we derived in the previous sections were general and apply even to the three-flavor neutrinos case. However, for the sake of clarity and simplicity, we shall apply now those results to the case of two flavors, and hence also to two mass eigenstates $j,k\in\{1,2\}$. In addition, we consider here only transitions among electron neutrinos $\nu_e$ and muon neutrinos $\nu_\mu$, so that $\alpha,\beta\in\{e,\mu\}$. The $\nu_e-\nu_\tau$ transition probabilities are computed in exactly the same way. This case is also the most relevant in view of its application to solar neutrinos. The mixing matrix $U^*_{\alpha j}$ in the two-flavor case has the following form in terms of the mixing angle $\theta$ \cite{NBook}:
\begin{equation}\label{UMatrix}
    U^*_{\alpha j}=
    \begin{pmatrix}
    \cos\theta & \sin\theta \\
    -\sin\theta & \cos\theta
    \end{pmatrix}.
\end{equation}
By plugging this matrix into Eq.\,(\ref{PlaneProbability}), we find the following transition probability in vacuum within the plane-wave approach for a given coordinate sparation $r_B-r_A$:
\begin{equation}\label{2FlavorPWProba}
    \mathcal{P}^{\rm PW}_{\nu_e\rightarrow\nu_\mu}= \tfrac{1}{2} \sin^22\theta\left[1-\cos\left(\frac{\Delta m_{21}^2\mathcal{J}}{2E_0}\right)\right].
\end{equation}
Here, the length $\mathcal{J}$ is the one given by the second integral in Eq.\,(\ref{ComputedPlanePhase}), and the algebraic oscillation length $L_{\rm osc}^{jk}$ as given by Eq.\,(\ref{PWOscLength}) has been replaced here by $L_{\rm osc}^{21}$. From this expression, the survival probability can also be computed using the identity, $\mathcal{P}^{\rm PW}_{\nu_e\rightarrow \nu_e}=1-\mathcal{P}^{\rm PW}_{\nu_e\rightarrow\nu_\mu}$. We would like to stress here that the validity of the latter identity for computing the survival probability holds only thanks to the conservation of unitarity under the conformal coupling as discussed in Secs.\,\ref{Sec:PW} and \ref{Sec:WP}.
On the other hand, plugging the matrix (\ref{UMatrix}) into Eq.\,(\ref{FinalConformalWPProba}), we find the following transition probability in vacuum within the wavepacket approach for the same coordinate distance $r_B-r_A=L$:
\begin{equation}\label{2FlavorWPProba}
    \mathcal{P}^{\rm WP}_{\nu_e\rightarrow\nu_\mu}=\tfrac{1}{2}\sin^22\theta\left[1-\cos\left(\frac{\Delta m_{21}^2\mathcal{J}}{2E_0}\right)\exp\left(-\left[\frac{L}{L^{21}_{\rm coh}}\right]^2\right)\right].
\end{equation}
The coherence length here is given by
\begin{equation}\label{2FlavorCoherence}
L_{\rm coh}^{21}\approx\frac{4\sqrt{2}E^2_0\sigma_x}{\Delta m_{21}^2}\left(1-\frac{m_2^2+m_1^2}{4E^2_0}\right).    
\end{equation}

For a concrete numerical example, we shall plot the variation of both probabilities with the distance traveled by the neutrinos within the chameleon model for which a detailed profile inside a constant-density and compact spherical object has been derived in Ref.\,\cite{ChameleonField}.  In what follows, we take $\Delta m^2_{12}=10^{-4}\,$eV$^2$, $E_0=0.5\,$MeV, $\sin^22\theta=1$ \cite{NeutrinoData} and $\sigma_x\sim10^{-8}\,$cm \cite{Coherence} for which we find $L_{\rm coh}^{21}\sim10^8\,$m. In addition, we take the scalar field's profile in vacuum outside the uniform-density star of radius $R_\odot$ (not necessarily the solar radius) to be given by the following equation \cite{ChameleonField}:
\begin{equation}\label{Profile}
    \phi(r)=\phi_\infty-K\frac{e^{-m_\infty R_\odot(x-1)}}{x}.
\end{equation}
Here, $\phi_\infty$ is the value of the scalar field outside the star, $m_\infty$ is the effective mass of the scalar field outside the star, $x=r/R_\odot$ so that $x\geq1$, and $K$ is a constant given by \cite{ChameleonField}:
\begin{equation}\label{K}
    K=\frac{\phi_cm_c^2R^2_\odot}{3(m_\infty R_\odot+1)}.
\end{equation}
The value of the scalar field inside the star is denoted here by $\phi_c$ and $m_c=\beta\rho_c/M_p\phi_c$ is related to the effective mass of the scalar field inside the star \cite{ChameleonField}. These formulas are valid approximations only when a constant density is assumed for the star inside of which the scalar field is assumed to vary radially, and $m_\infty R_\odot\ll1$ \cite{ChameleonField}. On the other hand, outside the star the spacetime metric is given by the Schwarzschild solution for which $\mathcal{A}(r)\mathcal{B}(r)=1$. Therefore, with the profile (\ref{Profile}) for the scalar field, integral $\mathcal{J}$ is computed as follows
\begin{align}\label{J-Integral}
    \mathcal{J}&=\int_{r_A}^{r_B}\Omega(\phi)\sqrt{\mathcal{A}(r)\mathcal{B}(r)}\,{\rm d}r\nonumber\\
    &=R_\odot\int_{1}^{x_B}\exp\left[\frac{\beta}{M_p}\left(\phi_\infty-K\frac{e^{-m_\infty R_\odot(x-1)}}{x}\right)\right]\,{\rm d}x\nonumber\\
    &\approx e^{\frac{\beta\phi_\infty}{M_P}}R_\odot \Bigg[\frac{r_B}{R_\odot}\exp\left(-\frac{\beta K}{M_p}\frac{R_\odot}{r_B}\right)-\exp\left(-\frac{\beta K}{M_p}\right)\nonumber\\
    &\qquad\qquad\qquad+\frac{\beta K}{M_p}{\rm Ei}\left(-\frac{\beta K}{M_p}\frac{R_\odot}{r_B}\right)-\frac{\beta K}{M_p}{\rm Ei}\left(-\frac{\beta K}{M_p}\right)\Bigg].
\end{align}
In the second line, we set $r_A=R_\odot$ and we denoted by $x_B$ the ratio $r_B/R_\odot$. In the last line, we assumed $m_\infty R_\odot\approx0$ \cite{ChameleonField}, and we introduced the exponential integral function ${\rm Ei}(x)=-\int_{-x}^\infty t^{-1}e^{-t}{\rm d}t$.

We now take the radius $R_\odot$ to be the radius of the Sun, the average density of the star to also be of the same order as that of the Sun, $\rho_c\sim1\,$g/cm$^3$, and the density of the environment outside the star to be that of the interplanetary medium, $\rho_\infty\sim10^{-24}\,$g/cm$^3$. Therefore, for a chameleon field with an $n=3$ power-law potential, we have $\phi_\infty=(\rho_c/\rho_\infty)^{\frac{1}{n+1}}\,\phi_c$ \cite{ChameleonField}, yielding $\phi_\infty\sim10^{6}\phi_c$. On the other hand, with the approximation $m_\infty R_\odot\approx0$, we have $K\sim10^4\phi_c$ for $m_c^2R_\odot^2=10^4$, for which the fifth force is highly suppressed compared to the Newtonian gravitational force outside the star \cite{ChameleonField}. Then, for $\phi_\infty\sim M_p$ and $\beta=1$ \cite{ChameleonIntroduction}, integral $\mathcal{J}$ approximates to
\begin{align}\label{NumericalJ-Integral}
    \mathcal{J}&\approx e \Bigg[(L+R_{\odot})\exp\left(-\frac{R_\odot}{100[L+R_\odot]}\right)-R_\odot\exp\left(-\frac{1}{100}\right)\nonumber\\
    &\qquad\qquad\qquad +\frac{R_\odot}{100}{\rm Ei}\left(-\frac{R_\odot}{100[L+R_\odot]}\right)-\frac{R_\odot}{100}{\rm Ei}\left(-\frac{1}{100}\right)\Bigg]\nonumber\\
    &\approx e(L+R_{\odot})\exp\left(-\frac{R_\odot}{100[L+R_\odot]}\right)-eR_\odot\exp\left(-\frac{1}{100}\right).
\end{align}
We have introduced here the coordinate distance $L=r_B-r_A$ and we have kept in the second step only the leading-order terms which allowed us to discard the difference between the last two terms inside the square brackets. This shorter expression is much easier to work with. Note that for values of $\phi_\infty$ much smaller than $M_p$, integral $\mathcal{J}$ would just reduce to the usual coordinate distance $L$, for which the transition probability would reduce to that of neutrinos not conformally coupled, and the effect of the scalar field would thus become totally screened. Therefore, the effect of the scalar field on neutrino flavor oscillations is to give rise to an {\it effective} path difference for neutrino mass eigenstates. The latter experience thus a longer path difference because of their interaction with the scalar field. Physically, this is very reminiscent of light rays propagating inside a refractive medium: two light rays experience a longer path difference when one of the rays propagates inside a medium of a higher refractive index. 

In fact, plugging our numerical values into the transition probability (\ref{2FlavorPWProba}) for a completely screened scalar field gives 
\begin{equation}\label{NumericalScreened2FlavorPWProba}
     \mathcal{P}^{\rm PW^0}_{\nu_e\rightarrow\nu_\mu}(L)\approx\frac{1}{2}-\frac{1}{2}\cos\left(\frac{5L}{10^4}\right),
\end{equation}
whereas plugging expression (\ref{NumericalJ-Integral}) into Eq.\,(\ref{2FlavorPWProba}) with the same numerical values used above, and then taking $R_\odot\sim10^9\,$m for simplicity\footnote{Note that considering a more compact object like a neutron star instead of the Sun has the advantage of providing us with a stronger gravitational field that would show up in a smaller radius $R_\odot$ and a higher central density $\rho_c$ entering our formulas (\ref{Profile}), (\ref{K}) and (\ref{J-Integral}). This would indeed remove the need to explore very large distances from the source to witness the effect. Fortunately, however, when considering the mean oscillation probability, as we do in Fig.\,3, our physical results become much clearer graphically even when restricting ourselves to the gravitational field of the Sun.}, gives
\begin{equation}\label{Numerical2FlavorPWProba}
\mathcal{P}^{\rm PW}_{\nu_e\rightarrow\nu_\mu}(L)\approx\frac{1}{2}\!-\!\frac{1}{2}\cos\Bigg[\left(\frac{L}{10^3}\!+\!10^6\right)\exp\left(-\frac{10^7}{L+10^9}\right)-\frac{10^6}{e^{0.01}}\Bigg],
\end{equation}
where $L$ displayed in these expressions is measured in meters. The variation of the two transition probabilities $\mathcal{P}^{\rm PW^0}_{\nu_e\rightarrow\nu_\mu}(L)$ and $\mathcal{P}^{\rm PW}_{\nu_e\rightarrow\nu_\mu}(L)$ with the distance $L$ is plotted in Fig\,.\,\ref{Fig:1} below.

The effect of the conformal coupling of neutrinos to the chameleon field is thus to modify indeed the effective traveled path of the neutrinos towards the detector as illustrated by the small spatial lag acquired by the thick line representing $\mathcal{P}_{\nu_e\rightarrow\nu_\mu}^{{\rm PW}}(L)$ with respect to the thin line representing $\mathcal{P}_{\nu_e\rightarrow\nu_\mu}^{{\rm PW}^0}(L)$ in Fig.\,\ref{Fig:1}. 
\begin{figure}
    \centering
    \includegraphics[scale=0.3]{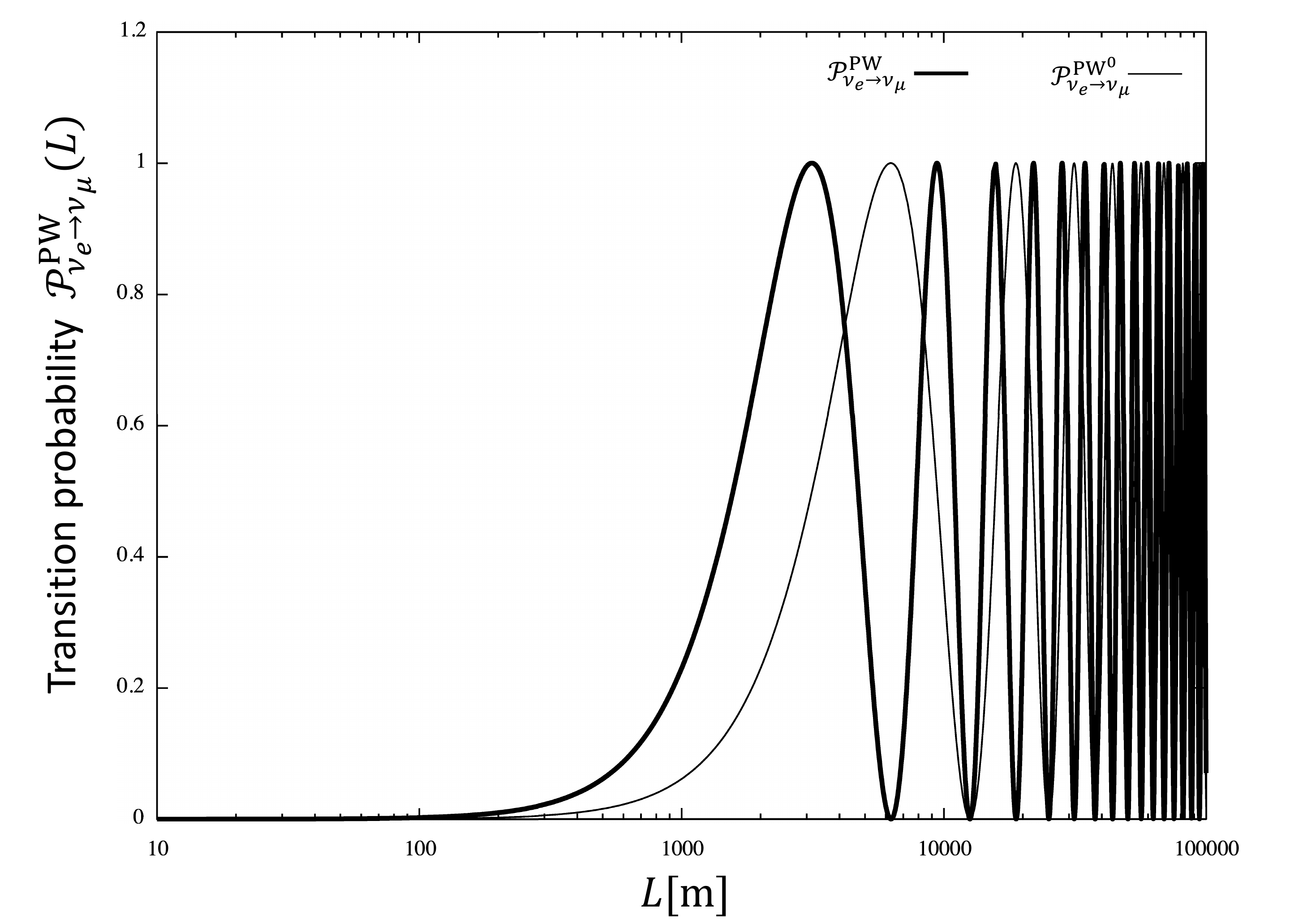}
    \caption{The variation (with the coordinate distance $L$) within the plane-wave approach of the transition probability $\mathcal{P}^{\rm PW}_{\nu_e\rightarrow\nu_\mu}(L)$ of conformally coupled neutrinos, as given by Eq.\,(\ref{Numerical2FlavorPWProba}), is represented here by the thick line. The thin line represents the variation of the transition probability  $\mathcal{P}^{\rm PW^0}_{\nu_e\rightarrow\nu_\mu}(L)$ as given by Eq.\,(\ref{NumericalScreened2FlavorPWProba}) for neutrinos not conformally coupled.}\label{Fig:1}
\end{figure}
\begin{figure}[H]
    \centering
    \includegraphics[scale=0.3]{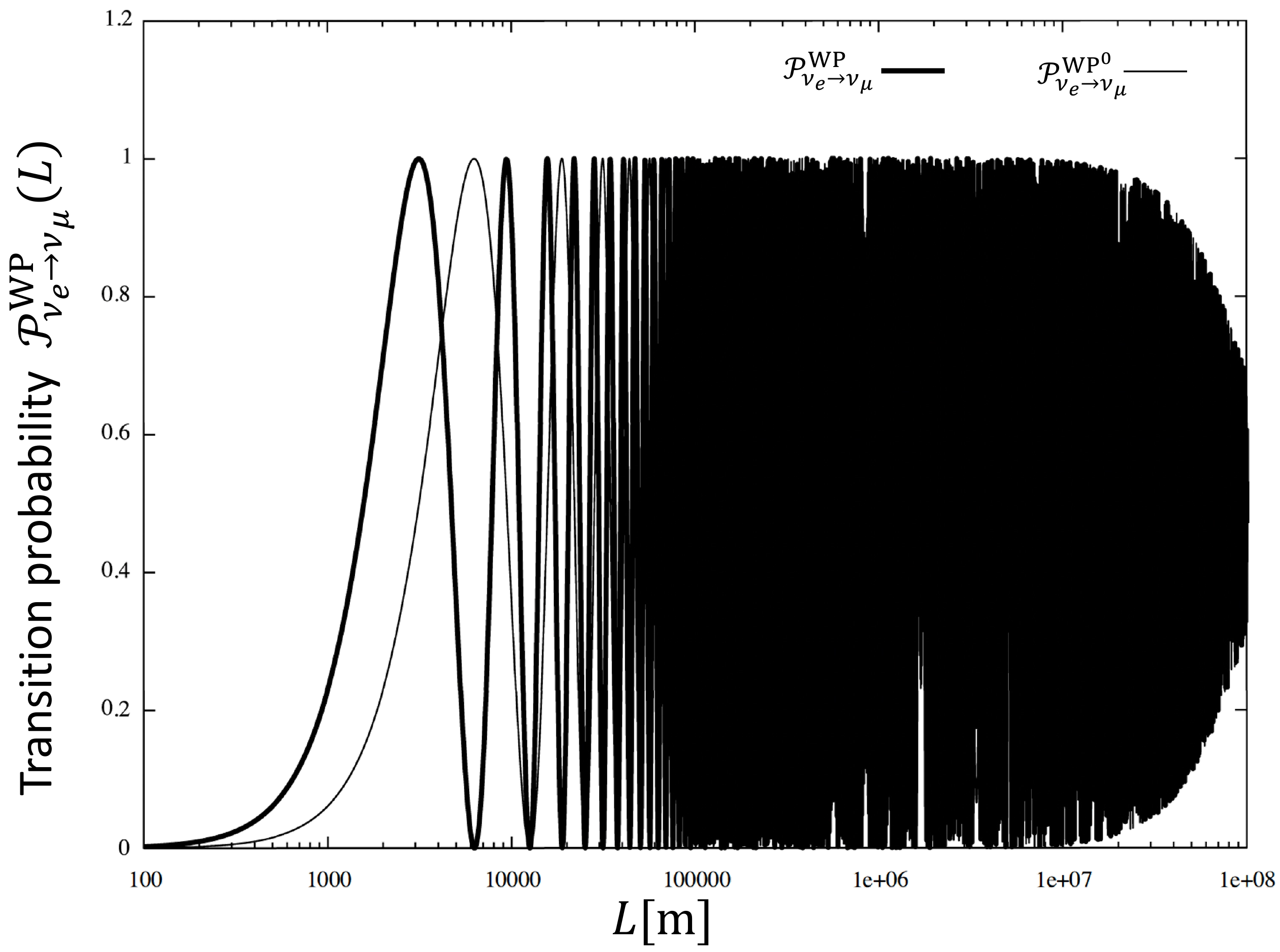}
    \caption{The variation (with the coordinate distance $L$) of the transition probability $\mathcal{P}^{\rm WP}_{\nu_e\rightarrow\nu_\mu}(L)$ given by expression (\ref{Numerical2FlavorWPProba}) within the wavepacket approach is represented by the thick line. The thin line represents the variation of the transition probability $\mathcal{P}^{{\rm WP}^0}_{\nu_e\rightarrow\nu_\mu}(L)$ of neutrinos not conformally coupled as given by Eq.\,(\ref{NumericalScreened2FlavorWPProba}).}.\label{Fig:2}
\end{figure}

Similarly, by plugging our numerical values into the transition probability (\ref{2FlavorWPProba}) for a completely screened scalar field within the wavepacket approach, we find
\begin{equation}\label{NumericalScreened2FlavorWPProba}
    \mathcal{P}^{\rm WP^0}_{\nu_e\rightarrow\nu_\mu}=\frac{1}{2}-\frac{1}{2}\cos\left(\frac{5L}{10^4}\right)\exp\left(-\frac{L^2}{10^{16}}\right),
\end{equation}
whereas plugging our numerical values as well as expression (\ref{NumericalJ-Integral}) for the $\mathcal{J}$ integral into Eq.\,(\ref{2FlavorWPProba}), gives the following expression for the transition probability of conformally coupled neutrinos within the wavepacket approach:
\begin{align}\label{Numerical2FlavorWPProba}
    &\mathcal{P}^{\rm WP}_{\nu_e\rightarrow\nu_\mu}(L)\approx\frac{1}{2}-\frac{1}{2}\exp\left(-\frac{L^2}{10^{16}}\right)\nonumber\\
    &\qquad\qquad\qquad \times\cos\Bigg[\left(\frac{L}{10^3}+10^6\right)\exp\left(-\frac{10^7}{L+10^9}\right)-\frac{10^6}{e^{0.01}}\Bigg].
\end{align}
The variation of the transition probabilities $\mathcal{P}^{\rm WP}_{\nu_e\rightarrow\nu_\mu}(L)$ and $\mathcal{P}^{\rm WP^0}_{\nu_e\rightarrow\nu_\mu}(L)$ with the distance $L$ are plotted in Fig\,.\,\ref{Fig:2}. The exponential damping displayed in the graphs is the same as the one experienced by neutrinos not conformally coupled, and is thus not affected by the conformal coupling of the neutrinos. Only the oscillation frequency is affected. Indeed, just like in the case of neutrinos not conformally coupled, we see that the transition probability tends to $\frac{1}{2}$ as the coordinate distance $L$ goes to infinity. This means that at long distances flavor oscillations become washed out by the damping due to the loss of coherence between the mass eigenstates' wavepackets. The probabilities for detecting either flavors become then equal. In other words, the transition probability at long distances tends to the value expected from a statistical mixture of the flavor states.

Now, because the transition probabilities (\ref{2FlavorPWProba}) and (\ref{2FlavorWPProba}) are highly oscillatory and it is not possible in practice to measure those oscillations for precise values of neutrino energies $E_0$ and traveled distances $L$, we shall extract from expressions (\ref{2FlavorPWProba}) and (\ref{2FlavorWPProba}) the oscillation probabilities averaged over a specific distribution of $L$ and $E_0$. For that purpose, we shall follow Ref.\,\cite{NBook} and assume, for simplicity, a normal distribution of $L/E_0$, with the mean value $\langle L/E_0\rangle$ and standard deviation $\sigma_{L/E_0}$. We assume $\sigma_{L/E_0}$ to be proportional to $\langle L/E_0\rangle$ and take for definiteness $\sigma_{L/E_0}=0.2\langle L/E_0\rangle$ as in Ref.\,\cite{NBook}.

Starting with Eq.\,(\ref{2FlavorPWProba}), we set $\mathcal{J}=L$ for the completely screened model and keep the same numerical values we chose previously, except that now we leave the energy $E_0$ as a variable measured in MeV. We then compute the average of the cosine function in that expression as follows:
\begin{align}\label{AvgCosine}
    \Biggl\langle\cos\left(\frac{2.5L}{10^4E_0}\right)\Biggr\rangle&\!=\!\!\int\!\!\cos\left(\frac{2.5L}{10^4E_0}\right)\frac{{\rm d}\left(\frac{L}{E_0}\right)}{\sigma_{L/E_0}\!\!\sqrt{2\pi}}\exp\left[-\frac{\left(\frac{L}{E_0}\!-\!\langle\frac{L}{E_0}\rangle\right)^2}{2\sigma^2_{L/E_0}}\right]\nonumber\\
    &=\cos\left(\frac{2.5}{10^4}\Biggl\langle\frac{L}{E_0}\Biggr\rangle\right)\exp\left(-\frac{2.5}{2\times10^{9}}\Biggl\langle\frac{L}{E_0}\Biggr\rangle^2\right).
\end{align}
Plugging this expression of the average cosine into Eq.\,(\ref{2FlavorPWProba}) gives the following mean oscillation probability for the completely screened scalar field:
\begin{equation}\label{AvgScreened2FlavorPWProba}
    \langle\mathcal{P}^{\rm PW^0}_{\nu_e\rightarrow\nu_\mu}\rangle= \frac{1}{2}-\frac{1}{2} \cos\left(\frac{2.5}{10^4}\Biggl\langle\frac{L}{E_0}\Biggr\rangle\right)\exp\left(-\frac{2.5}{2\times10^{9}}\Biggl\langle\frac{L}{E_0}\Biggr\rangle^2\right).
\end{equation}
On the other hand, to extract the mean oscillation probability for neutrinos coupled to the chameleon field we first expand $\mathcal{J}$ as given by Eq.\,(\ref{NumericalJ-Integral}) up to the first order in $L$ and get
$\mathcal{J}\approx eL$. Therefore, the average oscillation probability in this case reads
\begin{equation}\label{Avg2FlavorPWProba}
    \langle\mathcal{P}^{\rm PW}_{\nu_e\rightarrow\nu_\mu}\rangle= \frac{1}{2}-\frac{1}{2} \cos\left(\frac{2.5e}{10^4}\Biggl\langle\frac{L}{E_0}\Biggr\rangle\right)\exp\left(-\frac{2.5e^2}{2\times10^{9}}\Biggl\langle\frac{L}{E_0}\Biggr\rangle^2\right).
\end{equation}
Both transition probabilities as functions of the mean ratio $\langle L/E_0\rangle$ are plotted in Fig.\,\ref{Fig:3} below. We much more clearly see from this plot the effect of the scalar field on the flavor oscillations. The spatial lag between the two oscillation probabilities can distinctly be read off from the graph. 
\begin{figure}[H]
    \centering
    \includegraphics[scale=0.3]{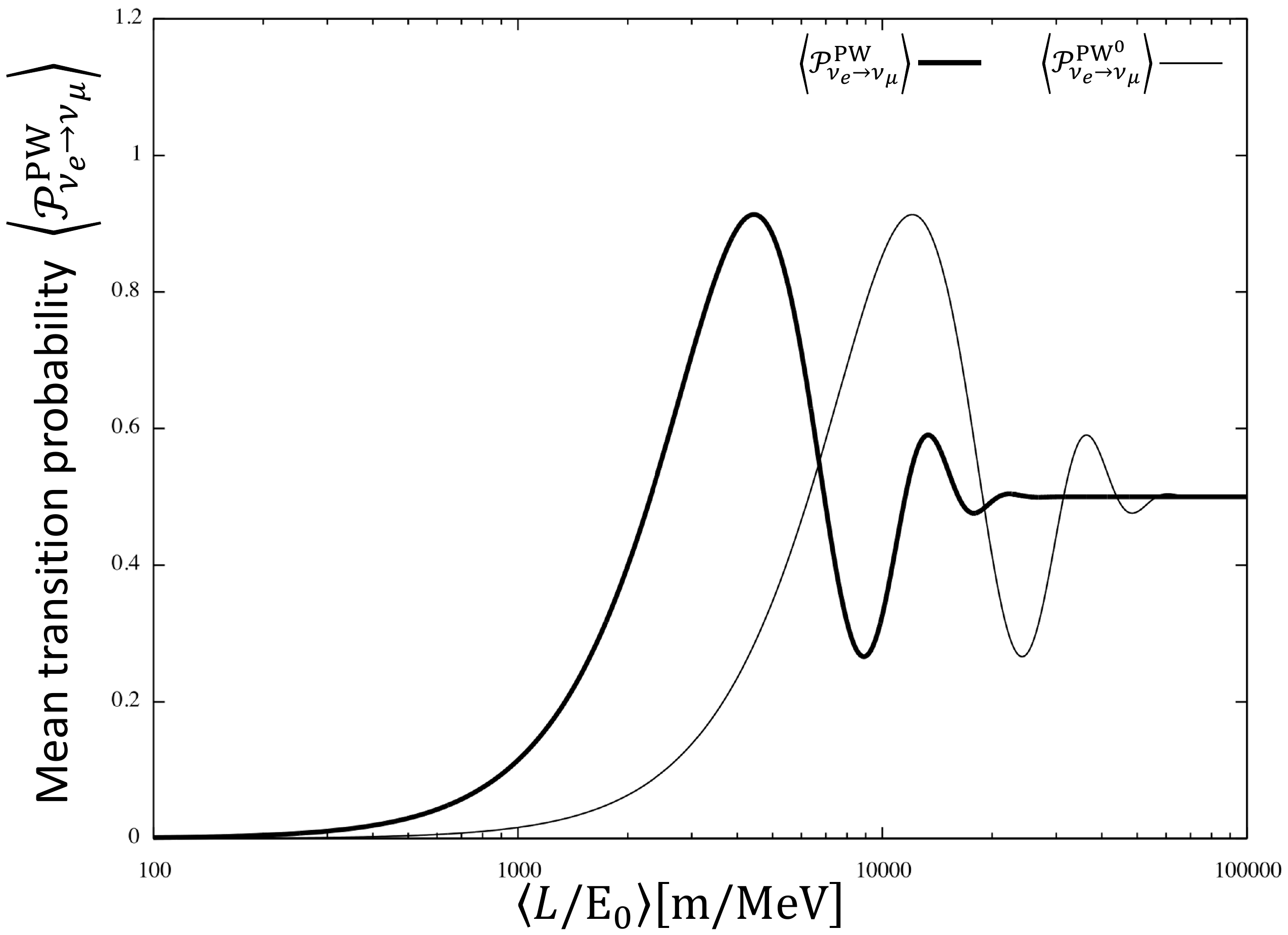}
    \caption{The variation (with the average ratio $\langle L/E_0\rangle$) of the mean flavor oscillation probability $\langle\mathcal{P}^{\rm PW}_{\nu_e\rightarrow\nu_\mu}\rangle$ given by expression (\ref{Avg2FlavorPWProba}) within the plane-wave approach is represented by the thick line. The thin line represents the variation of the mean flavor oscillation probability $\langle\mathcal{P}^{{\rm PW}^0}_{\nu_e\rightarrow\nu_\mu}\rangle$ of neutrinos not conformally coupled, as given by Eq.\,(\ref{AvgScreened2FlavorPWProba}).}\label{Fig:3}
\end{figure}
\section{The case of non-universal conformal couplings}\label{Sec:AppB}
In this section we describe how our previous results get modified when the conformal coupling of neutrinos is not universal. We say that the conformal coupling is non-universal when the scalar field $\phi$ and/or the conformal factor $\Omega(\phi)$ is different for different mass eigenstates of the neutrinos. We denote here the two possibilities at once by simply adding a subscript $j$; denoting thus by $\Omega_j(\phi)$ the conformal factor that interacts with the $j$-th mass eigenstate. 

\subsection{Plane-wave neutrinos}
When we dealt with neutrino plane waves in Sec.\,\ref{Sec:PW}, we had the option to either integrate independently over time and space, or to turn the time integral into a spatial integral by taking care of assigning to all the mass eigenstates the same arrival time $\mathcal{T}_*$. We argued that although the first option is not as rigorous as the second, it does nevertheless lead to the same final transition probability as the second. However, when allowing for a non-universal coupling we immediately see that the first option becomes totally inapplicable. In fact, the first option would assign to any $j$-th mass eigenstate a quantum phase that has the form 
\begin{align}\label{NonUniversalPWPhase}
\tilde{\Phi}_{j}^{\rm PW}&=-E_j(p)\int_{A}^{B}\Omega_j(\phi)\,{\rm d}t+E_j(p)\int_{r_A}^{r_B}\Omega_j(\phi)\sqrt{\frac{\mathcal{B}(r)}{\mathcal{A}(r)}}\,{\rm d}r\nonumber\\
&\quad-\frac{m_j^2}{2E_j(p)}\int_{r_A}^{r_B}\Omega_j(\phi)\sqrt{\mathcal{A}(r)\mathcal{B}(r)}\,{\rm d}r\nonumber\\
&\equiv-E_j(p)\,\mathcal{T}_j+E_j(p)\,\mathcal{I}_{j}-\frac{m_j^2}{2E_j(p)}\,\mathcal{J}_{j}.
\end{align}
It is clear that when substituting this phase into the expression (\ref{ConformalPlaneProba}) for the transition probability, no single time $\mathcal{T}$ would be available to be integrated over to extract an average transition probability.

If, on the other hand, we use the second option then it becomes possible to compute the transition probability by using again the common time given by Eq.\,(\ref{MassiveTime}). However, such a time, corresponding to the most massive eigenstate, takes now the following form:
\begin{align}\label{AppMassiveTime}
    \mathcal{T}_*&=\int_A^B\Omega_*(\phi){\rm d}t\nonumber\\
    &=\int_{r_A}^{r_B}\Omega_*(\phi)\sqrt{\frac{\mathcal{B}(r)}{\mathcal{A}(r)}}{\rm d}r+\int_{r_A}^{r_B}\frac{\tilde{m}_*^2\Omega_*(\phi)}{2\tilde{E}_*^2(\tilde{p})}\sqrt{\mathcal{B}(r)\mathcal{A}(r)}{\rm d}r\nonumber\\
    &\equiv\mathcal{I}_{*}+\frac{m_*^2}{2E_*^2(p)}\mathcal{J}_{*}.
\end{align}
Replacing the time integral $\mathcal{T}_j$ by this time $\mathcal{T}_*$ in the calculation (\ref{AppMassiveTime}) of phases, the resulting quantum mechanical phase a $j$-th mass eigenstate acquires is given by the following expression:
\begin{equation}\label{Sec6:trPhase}
    \tilde{\Phi}_j^{\rm PW}=E_j(p)\left(\mathcal{I}_{j}-\mathcal{I}_{*}\right)-\frac{m_j^2}{2E_j(p)}\,\mathcal{J}_{j}-\frac{E_j(p)m_*^2}{2E_*^2(p)}\,\mathcal{J}_{*}.
\end{equation}
Therefore, inserting into expression (\ref{ConformalPlaneProba}) of the transition probability the phase difference $\tilde{\Phi}^{\rm PW}_{jk}$ one obtains by subtracting from Eq.\,(\ref{Sec6:trPhase}) the corresponding expression for the $k$-th mass eigenstate, and then using $E_{j}(p)\approx E_0+\xi m_j^2/2E_0$ and $E_{k}(p)\approx E_0+\xi m_k^2/2E_0$, the transition probability takes the following form
\begin{multline}\label{Sec6:MassiveBasedProbability}
\mathcal{P}_{\alpha\rightarrow\beta}=\sum_{j,k}U^*_{\alpha j}U_{\beta j}U_{\alpha k} U^*_{\beta k}\exp\Bigg(-\frac{2\pi i}{L_{\rm osc}^{jk}}\Bigg[\frac{m_j^2\mathcal{J}_{j}-m_k^2\mathcal{J}_{k}}{\Delta m_{jk}^2}\\
-\frac{2E_0\Delta\mathcal{I}_{jk}}{\Delta m_{jk}^2}+\xi\left(\mathcal{I}_{*}-\frac{m_j^2\mathcal{I}_{j}-m_k^2\mathcal{I}_{k}}{\Delta m_{jk}^2}\right)\Bigg]\Bigg).
\end{multline}
We introduced here the notation $\Delta \mathcal{I}_{jk}\equiv\mathcal{I}_{j}-\mathcal{I}_{k}$. The oscillation length $L_{\rm osc}^{jk}$ is still as given by Eq.\,(\ref{PWOscLength}). Note that we have assumed here that the mixing matrix $U_{\alpha j}^*$ is the same for conformally coupled neutrinos even in the non-universal conformal coupling case (see Appendix \ref{App} for the case where this assumption is relaxed). In contrast to the universal coupling scenario, however, we see that the parameter $\xi$ appeared here explicitly in the transition probability even though we limited ourselves to the first order in $m_j^2/E_0$. 

From the result (\ref{Sec6:MassiveBasedProbability}) we find the following probability for an electron neutrino to be detected as a muon neutrino in the simple case of two-flavor neutrinos:
\begin{multline}\label{Sec6:2FlavorPWProba}
    \mathcal{P}^{\rm PW}_{\nu_e\rightarrow\nu_\mu}=\sin^22\theta\sin^2\Bigg(\frac{\Delta m_{21}^2}{4E_0}\Bigg[\frac{m_2^2\mathcal{J}_{2}-m_1^2\mathcal{J}_{1}}{\Delta m_{21}^2}\\
    +\xi\left(\mathcal{I}_{*}-\frac{m_2^2\mathcal{I}_{2}-m_1^2\mathcal{I}_{1}}{\Delta m_{21}^2}\right)\Bigg]-\frac{E_0\Delta\mathcal{I}_{21}}{2}\Bigg).
\end{multline}
This result neatly shows how a non-universal conformal coupling would affect the transition probability. It manifests itself in the difference between the ways each mass eigenstate experiences the spacetime path along which it travels. Because of the very last term inside the parentheses, we see that the correction becomes more significant for high-energy neutrinos. When spacetime paths are identically experienced by all the mass eigenstates, i.e., when $\mathcal{J}_2=\mathcal{J}_1$ and $\mathcal{I}_2=\mathcal{I}_1=\mathcal{I}_*$, we recover the result (\ref{2FlavorPWProba}) of the universal coupling scenario.
\subsection{Wavepacket neutrinos}
When dealing with neutrino wavepackets in the non-universal conformal coupling, the calculation proceeds along the same steps that led us to Eq.\,(\ref{Final2DCurvedWP}) and then to Eq.\,(\ref{ConformalWPProba}). We compute here the accumulated quantum mechanical phase of the wavepacket (\ref{Final1DCurvedWP}) by using the same recipe we used for computing the phase (\ref{Sec6:trPhase}) of each mass eigenstate. The result has then the same form as in Eq.\,(\ref{Sec6:trPhase}), except that $E_j(p)$ is replaced by the average energy $\bar{E}_j$ and $E_*(p)$ is replaced by the average energy $\bar{E}_*$ of the wavepacket of the most massive state. Consequently, the expression of the wavepacket of the $j$-th mass eigenstate at the detection point $r=r_B$, such that $r_B-r_A=L$, takes the following form
\begin{multline}\label{Sec6:Final2DCurvedWP}
    \tilde{\psi}_j(T,L)=\left(\frac{2\sigma_{p}^2}{\pi}\right)^{\frac{1}{4}}\exp\Bigg[-i\,\Bigg(\frac{m_j^2}{2\bar{E}_j}\,\mathcal{J}_{j}-\bar{E}_j\left(\mathcal{I}_{j}-\mathcal{I}_{*}\right)\\
    +\frac{\bar{E}_jm_*^2}{2\bar{E}_*^2}\,\mathcal{J}_{*}\Bigg)-\sigma_{p}^2(L-\varv_jT)^2\Bigg].
\end{multline}
Therefore, adopting again the first-order approximation for the average energies $\bar{E}_j=E_0+\xi m_j^2/E_0$, the transition probability takes, up to the first order in $m^2_j/E_0$, the following form
\begin{multline}\label{Sec6:ConformalWPProba}
\mathcal{P}_{\alpha\rightarrow\beta}(T,L)=\sum_{j,k}\,U^*_{\alpha j}U_{\beta j}U_{\alpha k} U^*_{\beta k}\left(\frac{2\sigma_p^2}{\pi}\right)^{\frac{1}{2}}\\
\times\exp\Bigg(-\frac{2\pi i}{L_{\rm osc}^{jk}}\Bigg[\frac{m_j^2\mathcal{J}_{j}-m_k^2\mathcal{J}_{k}}{\Delta m_{jk}^2}-\frac{2E_0\Delta\mathcal{I}_{jk}}{\Delta m_{jk}^2}\\
+\xi\left(\mathcal{I}_{*}-\frac{m_j^2\mathcal{I}_{j}-m_k^2\mathcal{I}_{k}}{\Delta m_{jk}^2}\right)\Bigg]-\sigma_{p}^2\left[\left(L-\varv_jT\right)^2+\left(L-\varv_kT\right)^2\right]\Bigg).
\end{multline}
Here, the oscillation length $L_{\rm osc}^{jk}$ is also given by the standard expression (\ref{PWOscLength}). Averaging now this probability over the unmeasured coordinate time $T$, we easily find the following final expression for the transition probability:
\begin{multline}\label{Sec6:FinalConformalWPProba}
\mathcal{P}_{\alpha\rightarrow\beta}(L)=\sum_{j,k}\,U^*_{\alpha j}U_{\beta j}U_{\alpha k} U^*_{\beta k}\\
\times\exp\Bigg(-\frac{2\pi i}{L_{\rm osc}^{jk}}\Bigg[\frac{m_j^2\mathcal{J}_{j}-m_k^2\mathcal{J}_{k}}{\Delta m_{jk}^2}-\frac{2E_0\Delta\mathcal{I}_{jk}}{\Delta m_{jk}^2}\\
+\xi\left(\mathcal{I}_{*}-\frac{m_j^2\mathcal{I}_{j}-m_k^2\mathcal{I}_{k}}{\Delta m_{jk}^2}\right)\Bigg]-\left[\frac{L}{L_{\rm coh}^{jk}}\right]^2\Bigg).
\end{multline}
The algebraic coherence length $L_{\rm coh}^{jk}$ is as given by Eq.\,(\ref{2FlavorCoherence}). Note that the non-universality of the conformal coupling does not affect the coherence length either.
From this result, we compute the transition probability for the two-flavor neutrinos case to be,
\begin{multline}\label{Sec6:2FlavorWPProba}
    \mathcal{P}^{\rm WP}_{\nu_e\rightarrow\nu_\mu}=\tfrac{1}{2}\sin^22\theta\,\Bigg[1-e^{-\frac{L^2}{(L_{\rm coh}^{jk})^2}}\cos\Bigg(\frac{\Delta m_{21}^2}{2E_0}\Bigg[\frac{m_2^2\mathcal{J}_{2}-m_1^2\mathcal{J}_{1}}{\Delta m_{21}^2}\\
    +\xi\left(\mathcal{I}_{*}-\frac{m_2^2\mathcal{I}_{2}-m_1^2\mathcal{I}_{1}}{\Delta m_{21}^2}\right)\Bigg]-E_0\Delta\mathcal{I}_{21}\Bigg)\Bigg].
\end{multline}
This result reduces, of course, to expression (\ref{2FlavorWPProba}) when the conformal coupling becomes universal.
\section{Conclusion and discussion}\label{Sec:Conclusion}
We have examined the effect a conformal coupling of neutrinos would have on their flavor oscillations when they are propagating in vacuum. We adopted a general static and spherically symmetric metric for the curved spacetime and derived the transition probabilities in vacuum both within the plane-wave approach and within the wavepacket approach. We showed that in this case there are two different ways of dealing with time in the plane-wave approach: Either by averaging the transition probability over the unmeasured time as usually done in the literature for neutrinos not conformally coupled, or by converting the time integral into a spatial integral and taking into account the difference of the times of arrival of the different mass eigenstates. We showed that both methods lead to the same correct oscillation length. However, we showed that the first method is not applicable when dealing with wavepackets of conformally coupled neutrinos. 

We have discussed in great detail the different possible ways of implementing the wavepacket approach and highlighted the additional subtleties the latter conceals due to the presence of the position-dependent scalar field $\phi$ along the path of the neutrinos. For simplicity, we studied only the case of a time-independent scalar field $\phi$ since the calculations would become unwieldy when the field $\phi$ depends on time. Then we examined in detail the possibility of implementing the approach that consists of evaluating the wavepacket at the detector and we showed that the issues it gives rise to are more serious than the ones that already arise in the absence of any conformal coupling. We numerically examined the simple case of two-flavor neutrinos and plotted the variation with distance of the transition probabilities within the chameleon model using both the plane-wave and wavepacket approaches.

Although screening models make it possible to violate the equivalence principle even when their scalar field couples universally to matter, we have nevertheless explored in this work the additional possibility of having a non-universal conformal coupling in those models. A non-universal coupling offers, indeed, an additional way of violating the equivalence principle. We derived the general formulas for the probability transitions for such a scenario. Our formulas showed that the violation of the equivalence principle through these flavor oscillations would be more important for high-energy neutrinos.

Finally, we would like to stress here the important fact that all the transition probabilities derived here never exceed unity, both with a universal and a non-universal coupling scenario. This preservation of unitarity is in fact what gives meaning to the flavor survival probability and allows one to compute the latter. We pointed out that this important conclusion is due to the fact that a conformal transformation on Dirac spinors preserves unitarity.



\section*{Acknowledgments}
The authors are grateful to the anonymous referees for their constructive comments that helped improve our manuscript. This work was supported by the Natural Sciences and Engineering Research Council of Canada (NSERC) Discovery Grant No. RGPIN-2017-05388; and by the Fonds de Recherche du Québec - Nature et Technologies (FRQNT). PS acknowledges support from Bishop's University via the Graduate Entrance Scholarship award.

\appendix
\section{The mixing matrix of conformally coupled neutrinos}\label{App}
\subsection{The universal conformal coupling case} 
We shall argue here that the unitary mixing matrix $U_{\alpha j}^*$ is not affected by the conformal coupling of neutrinos when such a coupling is universal. For that purpose, we shall work for simplicity with the case of two flavors and two mass eigenstates. The argument for the more general case of three flavors and three mass eigenstates proceeds in the same manner.

First, let us use the relation $\ket{\nu_\alpha}=\sum_jU_{\alpha j}^*\ket{\nu_j}$ and the form (\ref{UMatrix}) of the mixing matrix to write explicitly in the Hilbert space each of the two flavors in terms of the two mass eigenstates:
\begin{equation}\label{System1}
\begin{aligned}
  \ket{\nu_e}&=\cos\theta\ket{\nu_1}+\sin\theta\ket{\nu_2}, \\
  \ket{\nu_\mu}&=-\sin\theta\ket{\nu_1}+\cos\theta\ket{\nu_2}.
\end{aligned}
\end{equation}
On the other hand, we know that the neutrinos' squared-mass matrix $\mathbb{M}^2$ is diagonalized by the mixing matrix and takes the following form in the mass eigenstates basis:
\begin{equation}\label{DiagonalM}
\mathbb{M}^2=m_1^2\ket{\nu_1}\bra{\nu_1}+m^2_2\ket{\nu_2}\bra{\nu_2}.
\end{equation}
Combining Eqs.\,(\ref{System1}) and (\ref{DiagonalM}), the effective squared-masses of the flavor states $\ket{\nu_e}$ and $\ket{\nu_\mu}$ read
\begin{equation}\label{System2}
\begin{aligned}
  m_e^2=\bra{\nu_e}\mathbb{M}^2\ket{\nu_e}&=m_1^2\cos^2\theta+m_2^2\sin^2\theta,\\
  m_\mu^2=\bra{\nu_\mu}\mathbb{M}^2\ket{\nu_\mu}&=m_1^2\sin^2\theta+m_2^2\cos^2\theta.
\end{aligned}
\end{equation}
From these two equations, one deduces that the single mixing matrix parameter $\theta$ is given in terms of the squared-masses by the relation
\begin{equation}\label{CosTheta}
\cos^2\theta=\frac{m_e^2-m_2^2}{m_1^2-m_2^2}.    
\end{equation}
Recalling now that {\it any} mass $\tilde{m}$ of a conformally coupled particle should be related to the mass $m$ of its non-conformally coupled counterpart by $\tilde{m}=\Omega(\phi)m$ \cite{What?,Fulton}, we deduce that the dimensionless mixing parameter $\tilde{\theta}$ corresponding to the conformally coupled neutrinos is given by
\begin{equation}\label{CosTheta}
\cos^2\tilde{\theta}=\frac{\tilde{m}_e^2-\tilde{m}_2^2}{\tilde{m}_1^2-\tilde{m}_2^2}=\frac{m_e^2-m_2^2}{m_1^2-m_2^2}=\cos^2\theta.
\end{equation}
We thus conclude that the parameter $\theta$ is indeed not affected by the conformal coupling of neutrinos in this case.

\subsection{The non-universal conformal coupling case}
For the case of the non-universal conformal coupling of neutrinos, the argument we used above does not apply. Indeed, for such a case we have $\tilde{m}_1=\Omega_1(\phi)m_1$, $\tilde{m}_2=\Omega_2(\phi)m_2$ and $\tilde{m}_e=\Omega_e(\phi)m_e$. The conformal factors $\Omega_1(\phi)$ and $\Omega_2(\phi)$ are fundamental ones dictated by the screening model, whereas the conformal factor $\Omega_e(\phi)$ is an {\it effective} one emerging from the mixing. The only argument we have in this case for the invariance of the mixing matrix is that the latter is made of dimensionless parameters representing only the weights of a superposition of states in the Hilbert space. Those parameters have nothing to do with spacetime and configuration space, and hence should not {\it a priori} be affected by the conformal coupling of the particles to spacetime.

Nevertheless, for the sake of completeness we shall display here the modified formulas for the transition probabilities one obtains in case the screening model is assumed to affect somehow even the quantum superposition of the mass eigenstates. For that purpose, we assume the elements $\tilde{U}_{\alpha j}^*$ of the mixing matrix of conformally coupled neutrinos to be related to those of non-conformally coupled ones by $\tilde{U}_{\alpha j}^*=\Omega_{jk}(\phi) U_{\alpha k}^*$, where the elements $\Omega_{jk}(\phi)$ belong to a $3\times3$ matrix that depends on the specific screening model one adopts. In order to preserve the unitarity of the mixing matrix, the matrix $\Omega_{jk}$ itself must be unitary, i.e., its elements must satisfy $\sum_j\Omega_{jl}^*\Omega_{jk}=\delta_{lk}$. Therefore, we see that the factors $U^*_{\alpha j}U_{\beta j}U_{\alpha k} U^*_{\beta k}$ involved in the general transition probabilities (\ref{Sec6:MassiveBasedProbability}) and (\ref{Sec6:FinalConformalWPProba}) should be replaced here by $\tilde{U}^*_{\alpha j}\tilde{U}_{\beta j}\tilde{U}_{\alpha k} \tilde{U}^*_{\beta k}$. When expressed in terms of the matrix elements $U_{\alpha j}^*$, the transition probabilities (\ref{Sec6:MassiveBasedProbability}) and (\ref{Sec6:FinalConformalWPProba}) take then the following forms, respectively:
\begin{multline}\label{App:MassiveBasedProbability}
\mathcal{P}_{\alpha\rightarrow\beta}=\sum_{j,k}\sum_{l,m,n,p}\Omega_{jl}\Omega_{jm}^*\Omega_{kn}^*\Omega_{kp}U^*_{\alpha l}U_{\beta m}U_{\alpha n} U^*_{\beta p}\\
\times\exp\Bigg(-\frac{2\pi i}{L_{\rm osc}^{jk}}\Bigg[\frac{m_j^2\mathcal{J}_{j}-m_k^2\mathcal{J}_{k}}{\Delta m_{jk}^2}-\frac{2E_0\Delta\mathcal{I}_{jk}}{\Delta m_{jk}^2}\\
+\xi\left(\mathcal{I}_{*}-\frac{m_j^2\mathcal{I}_{j}-m_k^2\mathcal{I}_{k}}{\Delta m_{jk}^2}\right)\Bigg]\Bigg).
\end{multline}

\begin{multline}\label{App:FinalConformalWPProba}
\mathcal{P}_{\alpha\rightarrow\beta}(L)=\sum_{j,k}\sum_{l,m,n,p}\Omega_{jl}\Omega_{jm}^*\Omega_{kn}^*\Omega_{kp}U^*_{\alpha l}U_{\beta m}U_{\alpha n} U^*_{\beta p}\\
\times\exp\Bigg(-\frac{2\pi i}{L_{\rm osc}^{jk}}\Bigg[\frac{m_j^2\mathcal{J}_{j}-m_k^2\mathcal{J}_{k}}{\Delta m_{jk}^2}-\frac{2E_0\Delta\mathcal{I}_{jk}}{\Delta m_{jk}^2}\\
+\xi\left(\mathcal{I}_{*}-\frac{m_j^2\mathcal{I}_{j}-m_k^2\mathcal{I}_{k}}{\Delta m_{jk}^2}\right)\Bigg]-\left[\frac{L}{L_{\rm coh}^{jk}}\right]^2\Bigg).
\end{multline}


\end{document}